\documentclass{iopjournal}

\usepackage[backend=bibtex]{biblatex}
\usepackage{amsmath,amssymb}
\usepackage{graphicx} 
\usepackage{xcolor}
\usepackage{xkeyval}
\usepackage{physics}
\usepackage{floatrow}
\usepackage{printlen}
\usepackage{hyperref} 
\usepackage{cleveref}
\usepackage{orcidlink}

\usepackage[colorinlistoftodos, prependcaption]{todonotes}

\begin{document}

\articletype{Paper} 

\title{The stochastic discrete nonlinear Schr\"odinger equation: \\microscopic derivation and finite-temperature phase transition}

\author{Mahdieh Ebrahimi$^{1,*}$\orcid{0000-0001-9503-3013}, Barbara Drossel$^1$\orcid{0000-0001-7115-6182} and Wolfram Just$^2$\orcid{0000-0002-5200-0761}}

\affil{$^1$Institute for Condensed Matter Physics, TU Darmstadt\\ Hochschulstrasse 6,
D-64289 Darmstadt, Germany}

\affil{$^2$Institute of Mathematics, University of Rostock\\ Ulmenstraße 69,
D-18057 Rostock, Germany}

\affil{$^*$Author to whom any correspondence should be addressed.}

\email{mahdieh.ebrahimi@pkm.tu-darmstadt.de, barbara.drossel@pkm.tu-darmstadt.de, wolfram.just@uni-rostock.de}


\begin{abstract}
We study a stochastic version of the
one-dimensional discrete nonlinear Schr{\"o}dinger equation
(DNSE), which is derived from first principles,
and thus possesses all the properties required by statistical mechanics,
such as detailed balance and the H-theorem. The stochastic version
shows disordered and localised dynamics, and
displays a corresponding phase transition
at a finite temperature value.
The phase transition can be captured in a quantitative way by a mean-field type
approach. The corresponding coarsening dynamics shows an unexpected dependence
on the noise strength, which is reminiscent of stochastic resonance.
The phase transition is linked with negative temperature phase transitions, which have been reported recently for the Hamiltonian dynamics of the DNSE.
Our approach gives a clue to how these negative temperature phase transitions
can be implemented in experimental setups, which are inevitably coupled to a
positive temperature heat bath.
\end{abstract}

\section{\label{sec:level1} Introduction}
The one-dimensional discrete nonlinear Schr\"odinger equation (DNSE) is a Hamiltonian system with two conserved quantities that exhibits intriguing features such as localised modes (breathers) and negative temperature states \cite{BaIuLiVu_PR21}. This non-integrable model was initially proposed to describe the dynamics of a group of anharmonic oscillators. Its primary aim was to explain nonlinear localisation phenomena, specifically breathers, i.e. spatially localised excitations that oscillate over time.
These phenomena have subsequently been observed in various nonlinear optical waveguide arrays \cite{1158804,b3727bba76384918a1c381ef76df660d}, Bose-Einstein condensation in optical lattices \cite{science.1062612, PhysRevLett.86.2353} and crystal lattice vibrations\cite{PhysRevLett.82.3288}. In this type of system with two conserved quantities, namely energy and particle number, condensation into discrete breathers and a negative-temperature regime occur because the system exhibits an upper bound on energy \cite{PhysRev.103.20,FLACH1998181},
causing entropy to decrease as energy approaches this upper bound.
The statistical mechanics analysis of the Discrete Nonlinear Schrödinger Equation (DNSE) for an isolated system was initially conducted by Rasmussen \emph{et al.} \cite{rasmussen_statistical_2000}. They showed that a transition from a disordered regime to a breather regime occurs at negative temperatures. Later, Rumpf, through a series of publications, identified the probability distribution of breathers and demonstrated that the entropy as a function of energy exhibits nonanalytic behaviour at the transition \cite{rumpf_transition_2008,RUMPF20092067}. Furthermore, Rumpf established that the focusing process, characterised by the emergence and amplification of peaks, is driven by entropy generation in the state of small amplitude waves resulting from the two conserved quantities of the system \cite{rumpf_growth_2007,rumpf_simple_2004,RuNe_PRL01}. The dynamics of this system produces localisation of energy within a small number of isolated high-amplitude structures.
In these publications, it was also argued that in the thermodynamic limit, states with several breathers will ultimately equilibrate to a single breather state, situated within a background of random fluctuations.  However, numerical studies of nonequilibrium situations have shown that the relaxation process is extremely slow \cite{PhysRevLett.134.097102, Ng_2009}.
These authors give two main reasons for the slowness to equilibrium \cite{PhysRevLett.122.084102, e19090445}, one is that a system with localised states coarsens very slowly (entropic effect)\cite{Iubini_2017,Iubini_2013}, and the other is because of the dynamic decoupling of the high amplitude breathers from the background fluctuations (dynamical effects)\cite{BaIuLiVu_PR21}.
Recent papers \cite{GrIuLiMa_JSM21,gradenigo2021condensation,giachello2025localization} report a phase transition in the condensation region of the negative-temperature regime of the DNSE, explaining the model's thermodynamics in the microcanonical ensemble using large-deviation techniques, where the nearest-neighbour interaction is neglected, and more recently \cite{giusfredi_mean-field_2025} by a mean-field theory that includes a simplified nearest-neighbour interaction.

In this paper, we consider the model from the point of view of a canonical ensemble.
Our formulation retains a parameter that measures the strength of the nonlinear term while the second constant of motion of the DNSE, the normalisation $N$, is taken to be fixed, $N=1$. Such a choice will turn out to be well-suited for
our purpose, but one should keep in mind that a thermodynamic limit of large system size may
differ from approaches where both the energy and the normalisation are considered as extensive variables. 
We set up a dissipative stochastic model, which follows by coupling the Hamiltonian of the
DNSE to a microscopic heat bath (see appendix \ref{sec:a} for the details).
Basic features of the resulting stochastic discrete nonlinear Schr\"odinger equation (SDNSE)
are then discussed in section \ref{sec:2}, such as symmetries and general parameter dependencies
(see also appendix \ref{sec:b} where the phase diagram of the related Hamiltonian
system is discussed). In particular, we show that a change of sign in the coefficient of the nonlinear term effectively 
corresponds to a change of sign in temperature. As we have obtained our stochastic description from a consistent microscopic Hamiltonian model, the SDNSE obeys detailed balance and ensures relaxation towards a constrained canonical equilibrium distribution, which remains well defined even in the negative 
temperature regime (see appendix \ref{sec:c} for explicit calculations).
Section \ref{sec:3} is devoted to a predominantly numerical study of the phase transition and its properties
(some details of the numerical integration scheme are contained in appendix \ref{sec:d}).
We begin with computing the dependence of the energy on the temperature, the so-called caloric
equation of state. The dependence on the system parameters can be conveniently captured
by a properly rescaled temperature variable. We find clear signatures of a first-order phase transition between a disordered and a localised phase. The transient nonequilibrium dynamics will display a 
non-monotonic dependence on the strength of the noise, a feature reminiscent of stochastic resonance. 
Finally, we provide in section \ref{sec:4} an analytic approach for the phase transition which
agrees quantitatively with the results of the simulations. We combine mean-field considerations
and exact analytic results for the caloric equation of state of the linear model
(see appendix \ref{sec:e}) to predict the temperature dependence of the energy in the disordered 
and in the localised phase, as well as the critical temperature itself. We also discuss how our findings relate to results
available in the literature.

\section{The stochastic model}\label{sec:2}

The deterministic discrete nonlinear Schr\"odinger equation (DNSE) is a paradigmatic
Hamiltonian system. The Hamiltonian and the equation of motion of the DNSE can be written as
\begin{equation}
\label{2.1}
H_S=\sum_{k=1}^L \left(2 |c_k|^2-c_k \bar{c}_{k-1} - c_k \bar{c}_{k+1} 
-\frac{\alpha}{2} |c_k|^4\right) \, 
\end{equation}
and
\begin{equation}
    \dot c_k(t) = i\frac{\partial H}{\partial \bar c_k}\, .
\end{equation}
The wave function amplitudes $c_k$ and their complex conjugates $\bar c_k$ form pairs of canonical conjugate coordinates. 
We impose periodic boundary conditions $c_k=c_{k+L}$ where $L$ denotes the system size. The strength of the nonlinear potential is governed by the real-valued quantity $\alpha$. In addition to the energy $E=H$, the equations of motion preserves also the normalization 
\begin{equation}\label{2.1a}
N=\sum_{k=1}^L | c_k|^2 \, . 
\end{equation}
Using $\alpha$
as the system parameter allows us to keep the normalisation fixed, and we chose here $N=1$. A translation between this setting and the  standardised
non-dimensional version of the DNSE (where $\alpha = -1$ and normalisation is the model parameter) can be found e.g. in \cite{EBRAHIMI2025134905}.
The phase diagram of this model, based on the microcanonical distribution,
is fairly well understood (see for instance \cite{rasmussen_statistical_2000, rumpf_transition_2008, BaIuLiVu_PR21}), and for keeping the presentation self-contained we cover the essential details in appendix \ref{sec:b}.

We introduce here a class of stochastic models that describe the coupling
of the Hamiltonian \eqref{2.1} to a heat bath with inverse temperature $1/(k_B T) = \beta$, where the normalization $N$ is still preserved (see e.g. \cite{BoDe_CMP99}, for early alternative stochastic models and
\cite{IuLeLiPo_JSM13, HaOl_SPDE23, IuPo_PRL25}
for more recent developments of the stochastic models related
to the DNSE). Such a class of models is given by
the stochastic differential equations (see appendix \ref{sec:a} for a derivation
of this model from first principles)
\begin{eqnarray}\label{2.2}
\dot{c}_k(t)=i \frac{\partial H_S}{\partial \bar{c}_k}-\beta \frac{\sigma^2}{2} 
c_k(t)
\left( \frac{\partial H_S}{\partial \bar{c}_k} \bar{c}_k(t)
-\frac{\partial H_S}{\partial c_k} c_k(t)\right) + i \sigma c_k(t)
\xi_k(t)
\end{eqnarray}
where $\xi_k(t)$ denote uncorrelated real-valued Gaussian white noise
\begin{equation}
\langle \xi_k(t) \xi_\ell(s)\rangle = \delta_{k \ell} \delta(t-s) \, .
\end{equation}
We use a Stratonovich version of the stochastic integral, so that we
can essentially handle the noise as an ordinary function, and no advanced stochastic calculus needs to be applied. The real valued parameter $\sigma$ governs the strength of the noise.
The damping constant $\beta \sigma^2/2$ is determined by the 
inverse temperature $\beta$ of the heat bath and the coupling $\sigma$ to the heat bath.
The form of the  damping is determined by the requirements of detailed balance.
As a consequence, as we will see in the next subsection, the dynamics of the stochastic differential equation has all the properties expected from a physical model, 
such as an H-theorem and relaxation towards a suitable canonical equilibrium distribution. 

\subsection{Properties of the equations of motion}\label{sec:2.1}

The stochastic model, eq.~\eqref{2.2} has the features that are required for a system that approaches thermodynamic equilibrium. These features do not
depend on the particular form of the Hamiltonian \eqref{2.1}, as long as the
Hamiltonian dynamics preserves the normalization $N$, that means 
\begin{equation}\label{2.1.1a}
    \{H_S,N\} = i\sum_k\left(\frac{\partial H_S}{\partial \bar c_k}\frac{\partial N}{\partial c_k} - \frac{\partial H_S}{\partial c_k}\frac{\partial N}{\partial \bar c_k}\right) =0\, .
\end{equation}

It is easy to see that the stochastic dynamics preserves the normalisation as well, since
\begin{eqnarray}\label{2.1.1}
\dot{N}&=& \sum_{k=1}^L \left(c_k \dot{\bar c}_k + \bar c_k \dot c_k\right)\nonumber\\
&=& \{H_S,N\} -\beta \frac{\sigma^2}{2} \sum_{k=1}^L
\left(|c_k|^2 - |c_k|^2\right) 
\left( \frac{\partial H_S}{\partial \bar{c}_k} \bar{c}_k
-\frac{\partial H_S}{\partial c_k} c_k\right) \nonumber\\
&& + i \sigma \sum_{k=1}^L \left( |c_k|^2-|c_k|^2\right) \xi_k(t) =0\, .
\end{eqnarray}

The deterministic part of eq.~\eqref{2.2} consists of a conservative contribution and
a damping term. The Hamiltonian $H_S$ provides a Lyapunov function of these
deterministic dissipative equations of motion, since
\begin{eqnarray}\label{2.1.2}
\dot{H}_S&=&\{H_S,H_S\} -\beta \frac{\sigma^2}{2}
\sum_{k=1}^L \left( c_k \frac{\partial H_S}{\partial c_k} - \bar{c}_k
\frac{\partial H_S}{\partial \bar{c}_k}\right)
\left( \frac{\partial H_S}{\partial \bar{c}_k} \bar{c}_k
-\frac{\partial H_S}{\partial c_k} c_k\right) \nonumber \\ 
&=& -\beta \frac{\sigma^2}{2} \sum_{k=1}^L
\left| \frac{\partial H_S}{\partial \bar{c}_k} \bar{c}_k
-\frac{\partial H_S}{\partial c_k} c_k \right|^2 \leq 0
\end{eqnarray}
if $\beta>0$. Equality holds only if each term in the sum individually vanishes and that
 happens at the stationary points determined by $\partial H_S/\partial {c}_k=0$
subject to the constraint $N=1$. Hence, the deterministic dynamics converge
for almost all initial conditions to the minima of the Hamiltonian $H_S$.

The stochastic model, eq.~\eqref{2.2}, obeys detailed balance as can be seen
by analysing the corresponding Fokker-Planck equation
(see appendix \ref{sec:c} for details). The analysis of the Fokker-Planck equation shows additionally that the stationary
distribution of the stochastic dynamics
is given by a canonical equilibrium constrained by the condition
$N=1$, that means
\begin{equation}\label{2.1.3}
\rho_\beta(\{c_k,\bar{c}_k\})= \frac{1}{Z_\beta} \exp(-\beta H_S) \delta(N-1)
\end{equation}
where the partition integral is given by
\begin{equation}\label{2.1.4}
Z_\beta = \int_{\mathbb{C}^L} 
\exp(-\beta H_S) \delta(N-1) \prod_{k=1}^L d(c_k,\bar{c}_k) \, .
\end{equation}
It is worth mentioning that the integral converges no matter whether the Hamiltonian itself admits a ground state since the condition $N=1$ confines integration to a 
compact domain. Therefore, the stationary distribution applies for positive as well as for negative values of the inverse temperature $\beta$. Furthermore, an H-theorem can be established 
(see appendix \ref{sec:c}
for details), so that in systems of finite size the dynamics converge to the stationary
state described by eq.~\eqref{2.1.4}. However, as usual, the corresponding relaxation timescales may depend severely on the system size.

\subsection{Symmetries and parameter dependencies}\label{sec:2.2}

From now onwards, we will consider eq.~\eqref{2.2} with the special choice
eq.~\eqref{2.1}, that means we will discuss the stochastic discrete
nonlinear Schr\"odinger equation (SDNSE)

\begin{eqnarray}\label{2.2.1}
\dot{c}_k(t)&=&i\left(2 c_k(t)-c_{k-1}(t)-c_{k+1}(t)-\alpha |c_k(t)|^2 c_k(t)\right)
\nonumber \\&&-\beta \frac{\sigma^2}{2} c_k(t)\left(
c_k(t) (\bar{c}_{k+1}(t) + \bar c_{k-1}(t))
-\bar c_k(t)(c_{k+1}(t) + c_{k-1}(t)) 
\right)\nonumber \\
&&+ i \sigma c_k(t) \xi_k(t)\, .
\end{eqnarray}

The relevant parameters of this
model are the system size $L$, the on-site potential strength $\alpha$, the
strength of the noise $\sigma$, and the inverse temperature $\beta$. We will first focus on symmetries in the parameter space. Parts of the subsequent analysis will be based on numerical simulations. For completeness
and for the convenience of the reader, we will outline some details of our
basic numerical algorithm in appendix \ref{sec:d} (see as well \cite{EBRAHIMI2025134905}).

While the microscopic
derivation required $\beta$ to be positive, the stochastic differential equation
\eqref{2.2.1} and its stationary density \eqref{2.1.3}
remain well defined even if $ \beta$ takes negative values. In such
cases, the damping constant becomes negative, so that energy will be pumped into
the system (cf. eq.~\eqref{2.1.2}). 
This is in line with physical intuition, as energy will
flow from a hot negative-temperature heat bath into the system. 
Thanks to the conservation law $N=1$, the motion remains confined to a
compact set and does not diverge. Furthermore, the model exhibits the symmetry $(\alpha,\beta) \mapsto (-\alpha,-\beta)$
with respect to these two parameters. That can be seen by employing a series of
coordinate transformations to eq.~\eqref{2.2.1}, namely
$c_k \mapsto \hat{c}_k=\exp(i 2 t) c_k$,
$\hat{c}_k \mapsto \hat{d}_k=(-1)^k \bar{\hat{c}}_k$,
$\hat{d}_k \mapsto d_k=\exp(-i 2 t) \hat{d}_k$,
so that both parameter settings, $(\alpha,\beta)$ and $(-\alpha,-\beta)$ are conjugate. This symmetry is also reflected in the phase diagram (see Figure \ref{figb1}) as the diagram is invariant under the transformation
$(\alpha,E)\mapsto (-\alpha, 4-E)$. Therefore, we can focus our investigations on parameter values $\beta>0$. A change in the sign of $\beta$ is equivalent to changing the sign of $\alpha$ and adjusting the energy scale, that means swapping between the focusing and
defocusing case of the DNSE. We just remark that the case $\alpha>0$, $\beta>0$ corresponds to
the negative temperature regime in the standardised DNSE studies 
\cite{rumpf_transition_2008, BaIuLiVu_PR21, EBRAHIMI2025134905},
while $\alpha<0$, $\beta>0$ corresponds to the positive temperature regime in those studies. 

\begin{figure}[h!]
\centering
\makebox[\linewidth]{%
  \includegraphics[width=8cm]{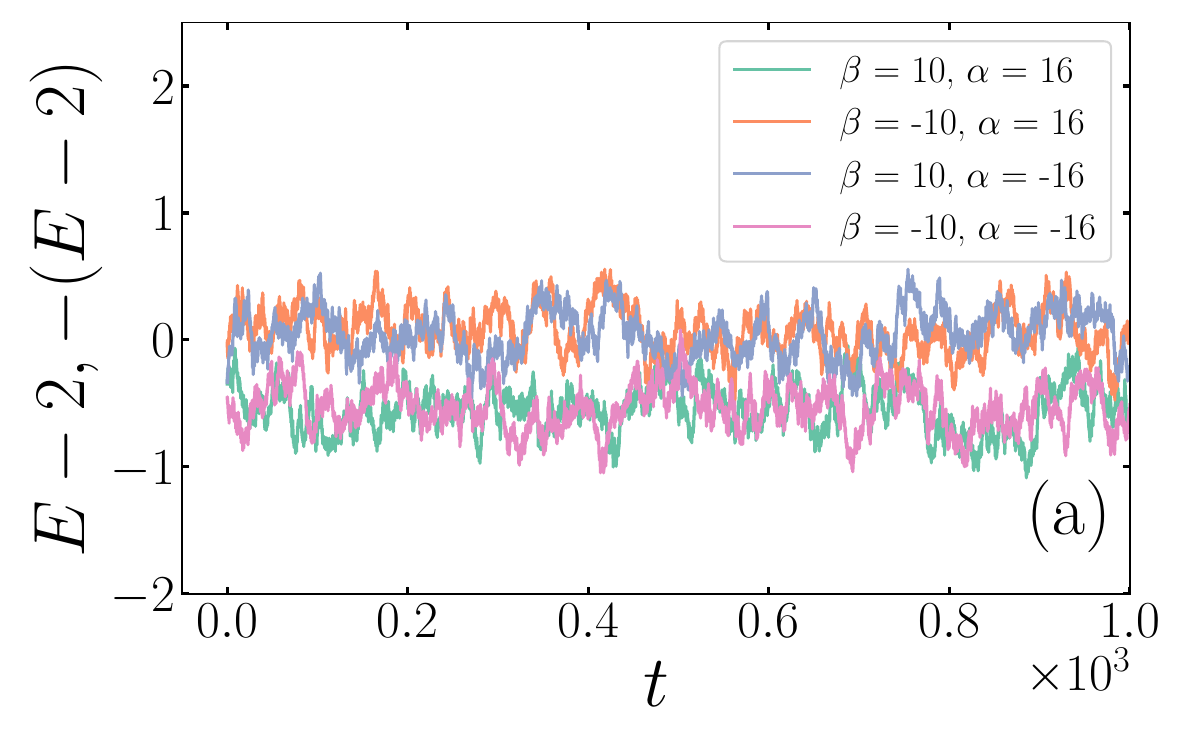} \hspace{0.1cm}
  \includegraphics[width=8cm]{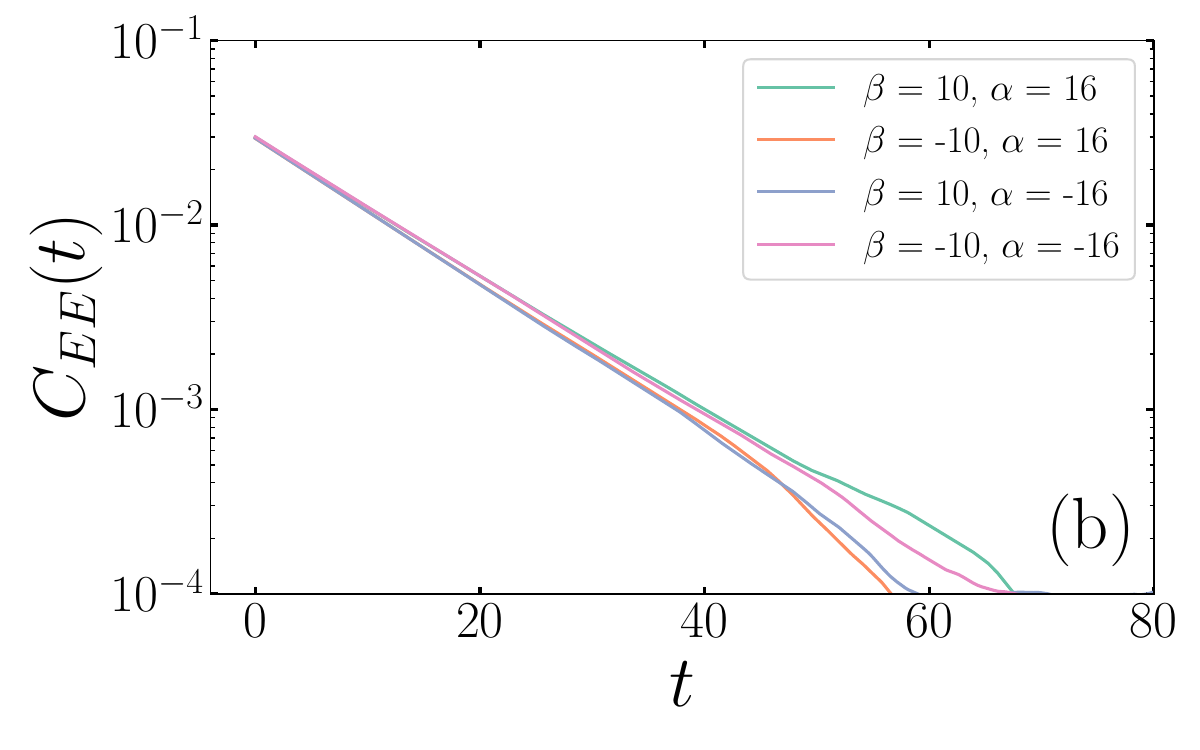}
}
\caption{
(a) Time traces of the energy $E(t)-2$ for parameter values $(\alpha,\beta)$ with positive $\alpha$ and
of $2-E(t)$ for conjugate parameter settings $(-\alpha,-\beta)$ obtained from numerical simulations of eq.~\eqref{2.2.1}  with system size $L=64$, noise strength $\sigma=0.3$ and stepsize
$\tau=0.01$, generated from a localised initial condition $c_n(0)=\delta_{n,32}$ by
skipping a transient of length about $\Delta t = 1.6 \times 10^5$ to ensure  stationary behaviour. Conjugate
parameter pairs: 
$(\alpha=16,\beta=10)$ (green), $(\alpha=-16, \beta=-10)$ (pink) and 
$(\alpha=-16,\beta=10)$ (blue), $(\alpha=16,\beta=-10)$ (orange).
(b) The associated energy autocorrelation functions on a semi-logarithmic scale obtained from a time series of length about
$T_{sim}=2.6 \times 10^{6}$, showing again the coincidence of data with conjugate parameter values. Parameter settings and colour coding as in part (a).}
\label{fig21}
\end{figure}

The aforementioned symmetries can be demonstrated by looking at numerical
simulations in the stationary regime. Figure \ref{fig21} shows time traces of the energy
$E(t)-2=H_S(t)-2$  and of $2-E(t)=2-H_S(t)$ for conjugate parameter settings, $(\alpha,\beta)$ and $(-\alpha,-\beta)$. The same symmetry can be seen when looking at the
autocorrelation function of the energy,
$C_{EE}(t)=\langle E(t) E(0) \rangle - \langle E(0)\rangle^2$,
where the average $\langle . \rangle$ is taken over a stationary ensemble of initial conditions $E(0)$, which can for instance be realised by a long time average. For practical purposes the autocorrelation function is computed via the Fourier transfrom of a stationary time series, using the Wiener Khinchin theorem.

\begin{figure}[h!]
\centering
\makebox[\linewidth]{%
  \includegraphics[width=8cm]{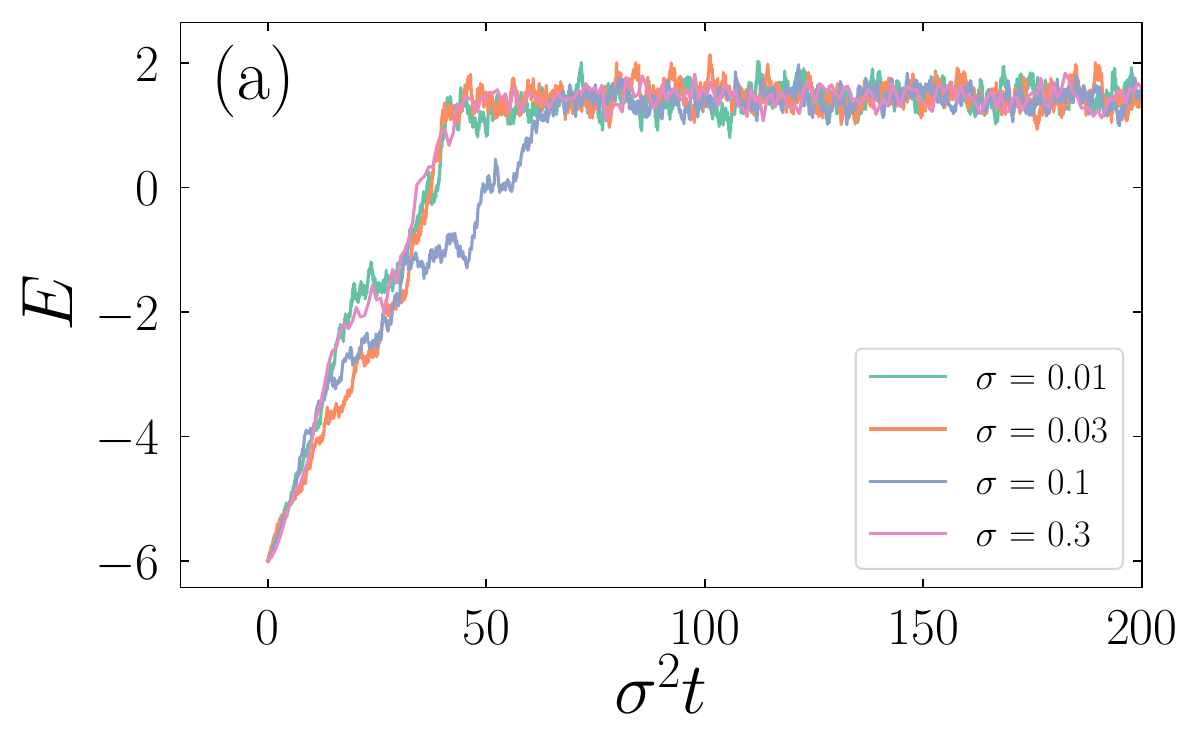} \hspace{0.1cm}
  \includegraphics[width=8cm]{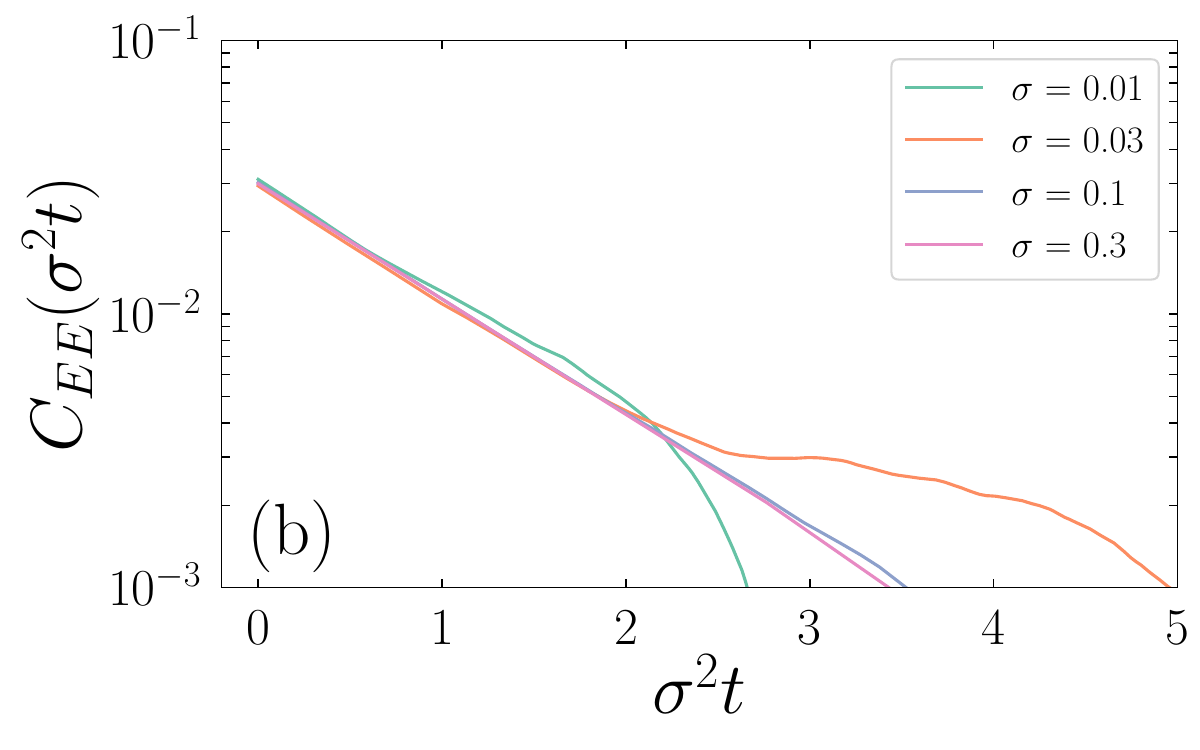}
} 
\caption{
(a) Time traces of the energy $E(t)$ for 
a system of size $L=64$ with parameter values $\alpha=16$, $\beta=10$ and different noise strength: $\sigma=0.01$ (green), $\sigma=0.03$ (orange), $\sigma=0.1$ (blue) and $\sigma = 0.3$ (pink). Data have been
obtained from a simulation of eq.~\eqref{2.2.1} with stepsize $\tau=0.01$ and initial 
condition $c_n(0)=\delta_{n,32}$. The energy is shown as a function of the rescaled time variable
$\sigma^2 t$. The data collapse shows that $\sigma^2 t$ sets the relevant time scale of the system. (b) Autocorrelation function of the energy 
in the stationary state, 
as a function of $\sigma^2 t$ for the parameter setting (and colour coding) of part (a).
Data have been computed from a time series of length about $T_{sim}=2.6 \times 10^{6}$ by skipping a transient of 
length about $\Delta t=16 \times 10^4$. The deviations from exponential decay for small values of $\sigma$ are due to sampling errors in the numerical computation.
}
\label{fig22}
\end{figure}

The timescales in the model \eqref{2.2.1} are determined, among others, by the internal dynamics of the Hamiltonian \eqref{2.1} as well as by the diffusion caused by noise. 
The noise strength $\sigma$ cannot be eliminated by a scaling of time. However,
we expect that increasing the strength of the noise speeds up the dynamics to
some extent (see as well \cite{EBRAHIMI2025134905} for analogous results in the infinite
temperature case $\beta=0$). By looking at the energy itself, $E(t)=H_S(t)$, we suspect seeing the sole diffusive effect of the noise, as the internal Hamiltonian
dynamics does not play a role in the time evolution of the energy. Figure \ref{fig22}
shows the transient dynamics of the energy starting from a localised initial
condition $c_n(0)=\delta_{n,32}$ as well as the autocorrelation of the energy in the
stationary state, displayed on the diffusive time scale $\sigma^2 t$. 
%
The observed data collapse confirms in these cases the simple picture that the noise strength predominantly determines the diffusive behaviour of the dynamics. 
Below, we will see  that for lower temperatures (i.e., larger $\beta$), where the system shows localised states, the impact of $\sigma$ on the dynamics becomes less trivial.

\section{Finite temperature phase transition} \label{sec:3}

The Hamiltonian dynamics of the DNSE have been extensively discussed in the context
of negative temperature states \cite{IuFrLiOpPo_NJP13} and related phase transition-like behaviour 
\cite{rasmussen_statistical_2000, RuNe_PRL01, MiKaDaFl_PRL18}. 
Recently, a phase transition line has been reported
\cite{GrMa_JSM19, GrIuLiMa_JSM21} which shows a nontrivial 
system size dependence. 
This phase transition was found in the negative temperature regime based on analytical calculations in the microcanonical ensemble and stochastic simulations that strictly preserve energy and normalisation. With our choice of a negative sign of the nonlinearity in the Hamiltonian eq.~\eqref{2.1}, such a phase transition is expected to occur at positive temperatures. This allows us to  uncover a phase
transition in a setup when the DNSE is coupled to a heat bath using our stochastic model.

\subsection{Caloric equation of state}\label{sec:3.1}

\begin{figure}[h!]
\centering
\makebox[\linewidth]{%
  \includegraphics[width=8cm]{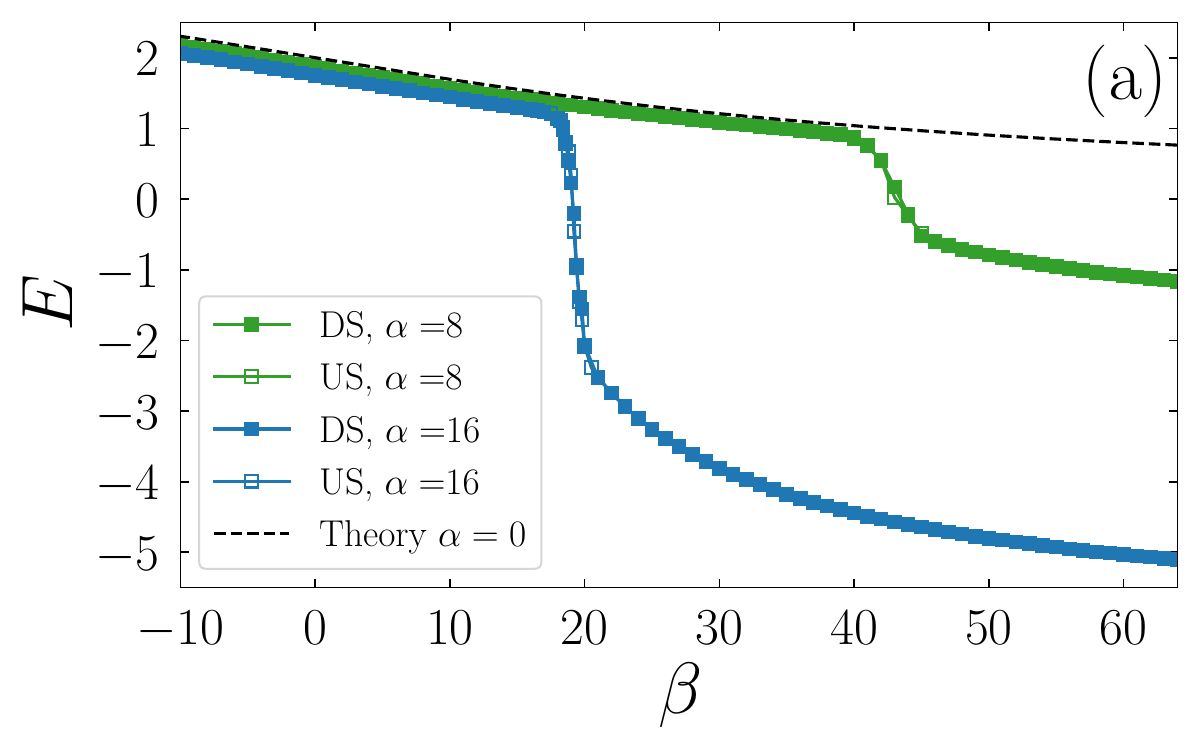} \hspace{0.1cm}
  \includegraphics[width=8cm]{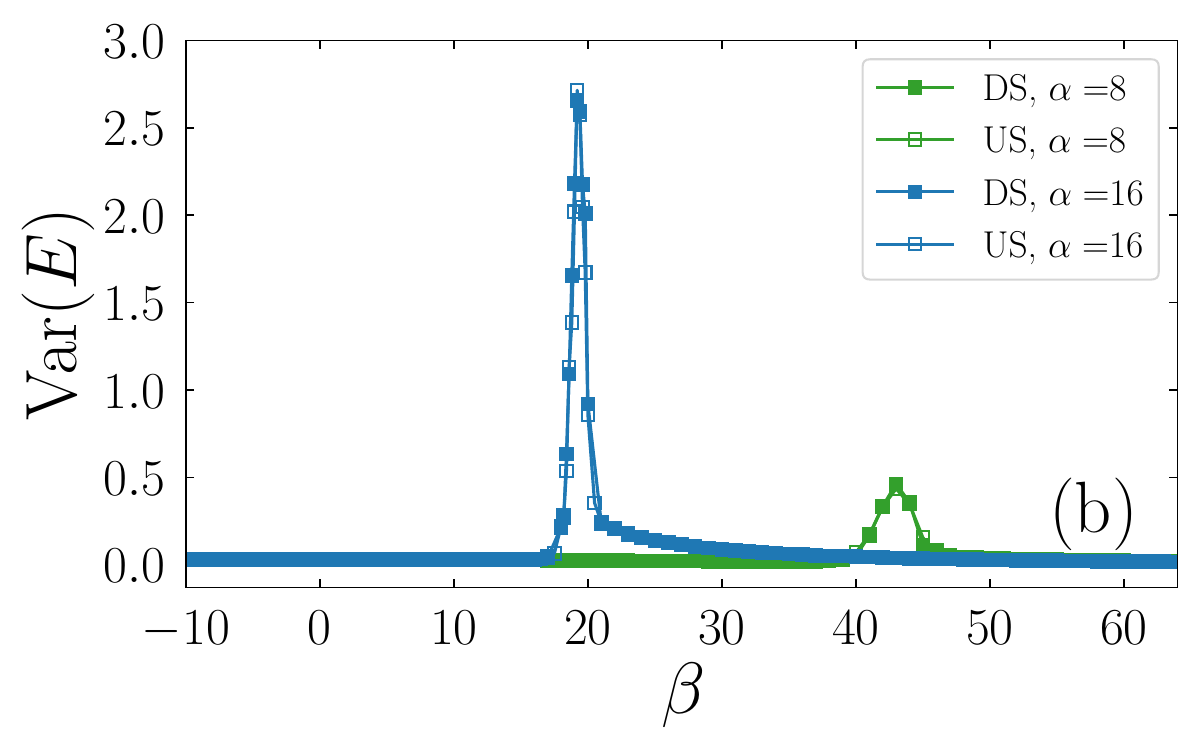}
} 
\caption{
(a) Mean energy $E(\beta)$ computed from time traces of eq.~\eqref{2.2.1} of length about $T_{sim}=2.6 \times 10^{6}$ (with longer time traces $T_{sim}=10^{7}$ in the transition region) skipping a transient of length about $\Delta t= 1.6 \times 10^{5}$ for system size $L=64$, noise strength $\sigma=0.3$, and two different values of $\alpha$: $\alpha=8$ (green) and $\alpha=16$ (blue). Computations were performed using an adiabatic parameter upsweep (open symbols) and downsweep (full symbols) of $\beta$; the data for upsweep and downsweep coincide perfectly. The black dashed line is the analytical result eq.~\eqref{e.16} obtained for the case $\alpha=0$. 
(b) Variance of the energy for the same parameter setup (and colour coding) as in part (a).}
\label{fig31}
\end{figure}

To supplement the basic phase diagram of the DNSE (see Figure \ref{figb1}) with temperature information, the dependence of energy on the temperature $E(\beta)$, i.e. the caloric equation of state, would be required. In principle, that information can be obtained from the known stationary distribution $\rho_\beta$, eq.~\eqref{2.1.3}, but as usual, the evaluation of 
phase space integrals becomes a challenge, to put it mildly. The linear case 
$\alpha=0$ can be dealt with in closed analytic form (see appendix \ref{sec:e}) and results in eq.~\eqref{e.16}, with an unusual dependence on the system 
size caused by the additional conservation law $N=1$. Thus, to discuss the caloric equation of state, we rely here, predominantly, on numerical computations of mean energies using time traces of eq.~\eqref{2.2.1} for $\alpha>0$. Figure \ref{fig31} shows the result of adiabatic temperature sweeps covering both negative and positive temperature regions. To initiate the parameter downsweep, a localised initial condition was used, $c_n(0)=\delta_{n,32}$,
while for the upsweep, a uniformly random initial condition was drawn from the distribution
$\delta(N-1)$ \footnote{For practical reasons, an almost normalised
state was drawn from the related canonical distribution $\exp(-N/L)$, with the normalisation
applied at the end.}. Both parameter sweeps yield almost identical mean values,
indicating the stationarity of the computation. A steep drop in the energy and large energy fluctuations indicate a phase transition at a positive $\beta$ value that depends on the value of $\alpha$. Such phase transition
characteristics are not visible for $\beta<0$. The dependence of the mean energy on $\beta$ in the low $\beta$-regime  hardly depends on the value of $\alpha$ and is quite well described by the
caloric equation of state of the harmonic chain, eq.~\eqref{e.16} (see the broken line in Figure \ref{fig31}).

\begin{figure}[h!]
\centering
\makebox[\linewidth]{%
  \includegraphics[width=8cm]{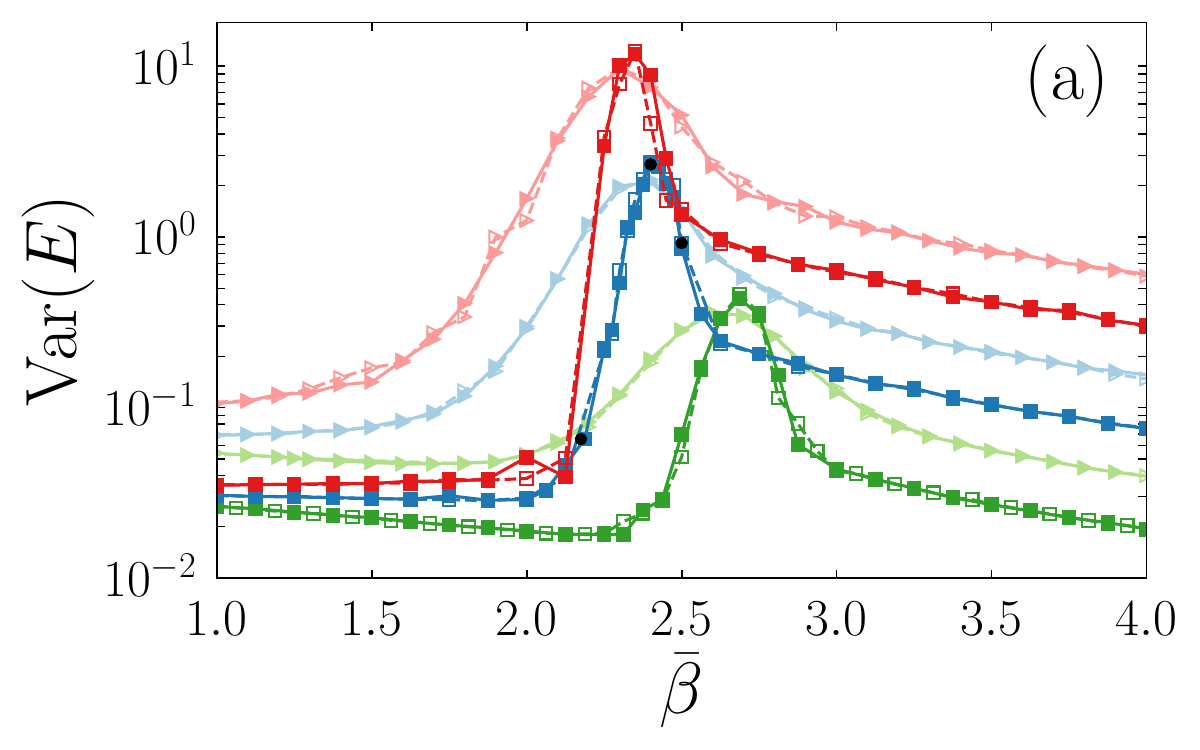} \hspace{0.1cm}
  \includegraphics[width=8cm]{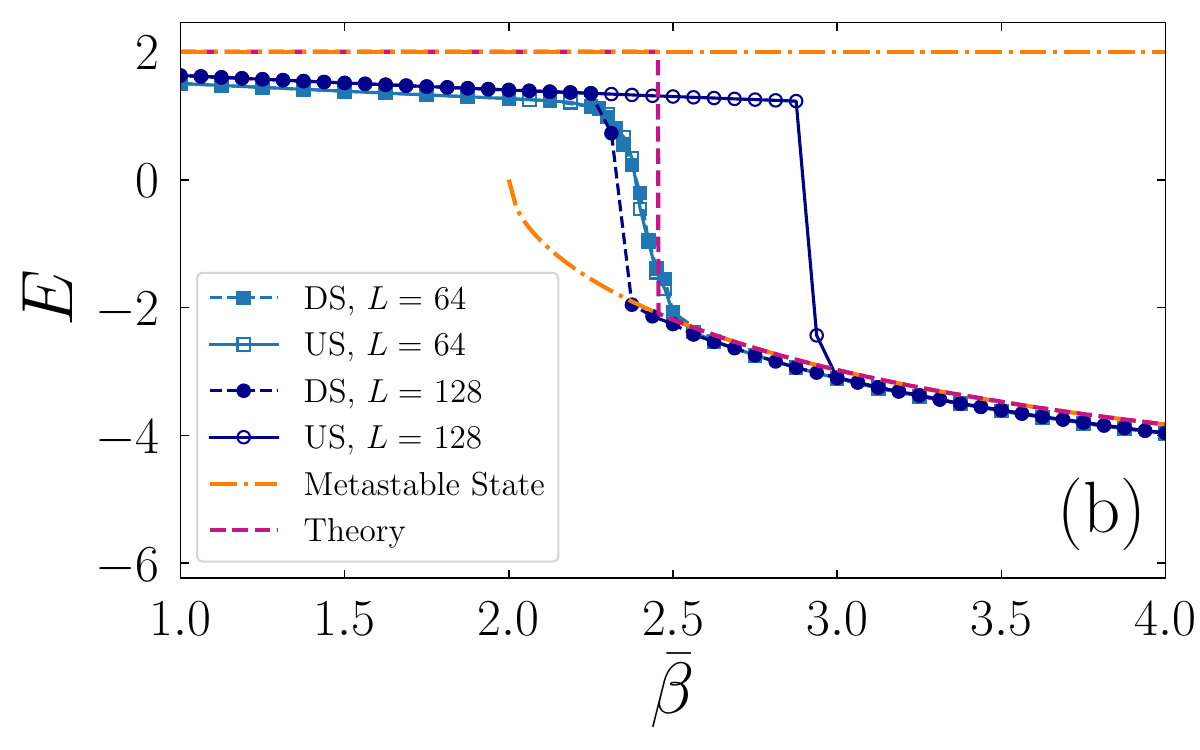}
}
\caption{
(a) Variance of the energy in dependence on the rescaled inverse temperature variable
$\bar{\beta}=\alpha \beta/(2 L)$ for $\sigma=0.3$ and $L=32$ (triangle symbols/lighter colours), $L=64$ (square
symbols/darker colours), $\alpha=8$ (green), $\alpha=16$ (blue), $\alpha=32$ (red), showing that the location of the phase transition depends essentially on the parameter $\bar\beta=\beta\alpha/2L$. Data have been computed by
adiabatic parameter upsweeps (solid) and downsweeps (dashed)
taking time averages over an interval of length about
$T_{sim}= 2.6 \times 10^{6}$ while skipping a transient of length about $\Delta t = 1.6 \times 10^{5}$. The three black points
indicate the particular temperature values used in Figure \ref{fig33}.
(b) Mean energy $E(\bar{\beta})$
for a system of size $L=128$ (dark blue circles) and $L=64$  (blue squares)
with $\alpha=16$ and $\sigma=0.3$, showing hysteresis due to extensive equilibration times for larger system sizes. 
Data have been computed by 
adiabatic parameter upsweeps (open symbols, solid) and downsweeps (full symbols, dashed)
taking time averages over an interval of length about
$T_{sim}=2.6 \times 10^{6}$ while skipping a transient of length about $\Delta t=1.6 \times 10^{4}$. The red dash line is the equilibrium energy of the system predicted by equation \eqref{4.1a}, with the orange dashed-dotted line indicating
analytic estimates of metastable states (see section \ref{sec:4} for details).}
\label{fig32}
\end{figure}

To obtain a better overview, we perform simulations for a larger range of parameter
settings. Inspired by the partition integral, eqs.~\eqref{2.1} and \eqref{2.1.4}, which depends on the product $\alpha \beta/2$ when we discount for the interaction between 
sites, and by the particular system size dependence shown in eq.~\eqref{e.16} it is tempting
to introduce the rescaled inverse temperature variable $\bar{\beta}=\alpha \beta/(2L)$. In fact,
with such a rescaled variable, the phase transition seems to appear around
$\bar{\beta}\approx 2.5$ for a larger set of parameters as
indicated by a pronounced peak in the variance of the energy, see Figure \ref{fig32}(a).
Below in section \ref{sec:4}, we will support this observation with a simple mean-field model that yields a phase transition around $\bar\beta = 2.45$.  
When increasing the system size, the phase transition is accompanied, as expected,
by large time scales. That results in numerical hysteresis, reminiscent
of first-order phase transitions, as shown in Figure
\ref{fig32}(b). 
Increasing the noise strength seems to accelerate the system dynamics 
as suggested by the results of the previous section
(for a more detailed discussion of the impact of noise we refer the reader to section \ref{sec:3.2})). Therefore increasing the noise strength
partially suppresses numerical hysteresis, see Figure \ref{fig32}(b).
Having said that, the time scales may 
become too long to be covered by standard numerical simulation times, and only quasi-stationary behaviour can be realised.

\subsection{Localised and disordered phase}\label{sec:3.2}

\begin{figure*}[ht!]
\centering
\makebox[\linewidth]{%
  \includegraphics[width=8cm]{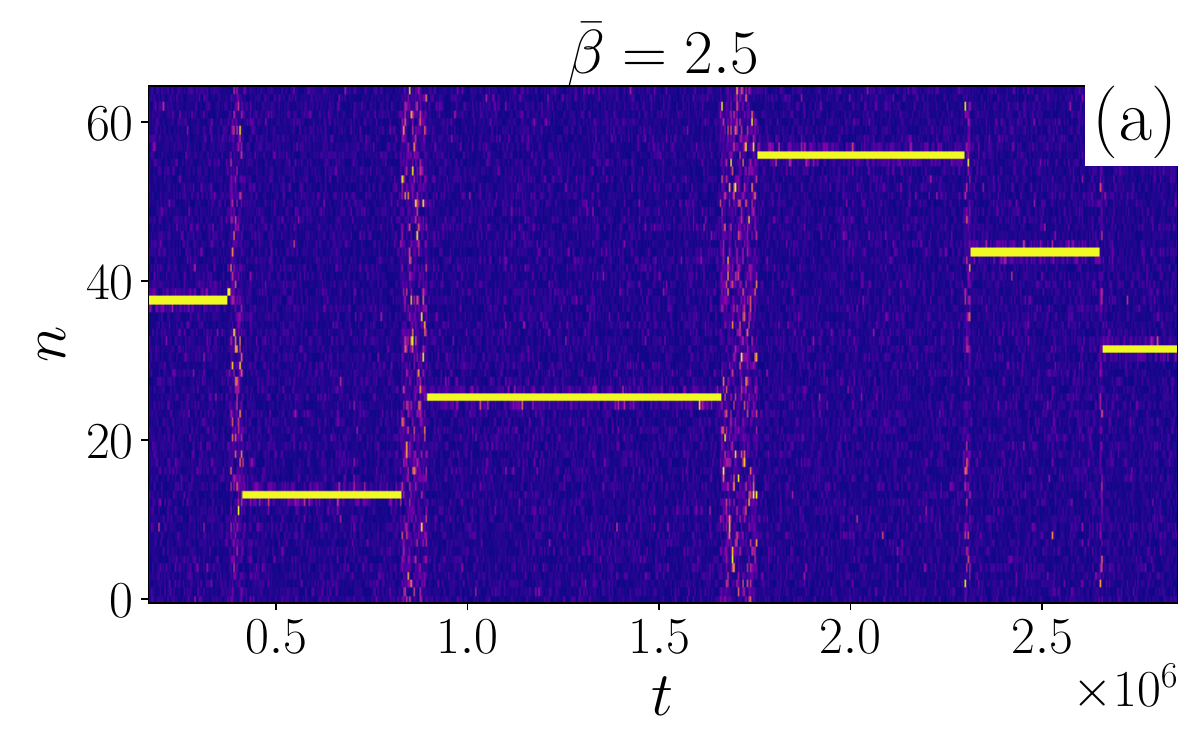} \hspace{0.1cm}
  \includegraphics[width=8cm]{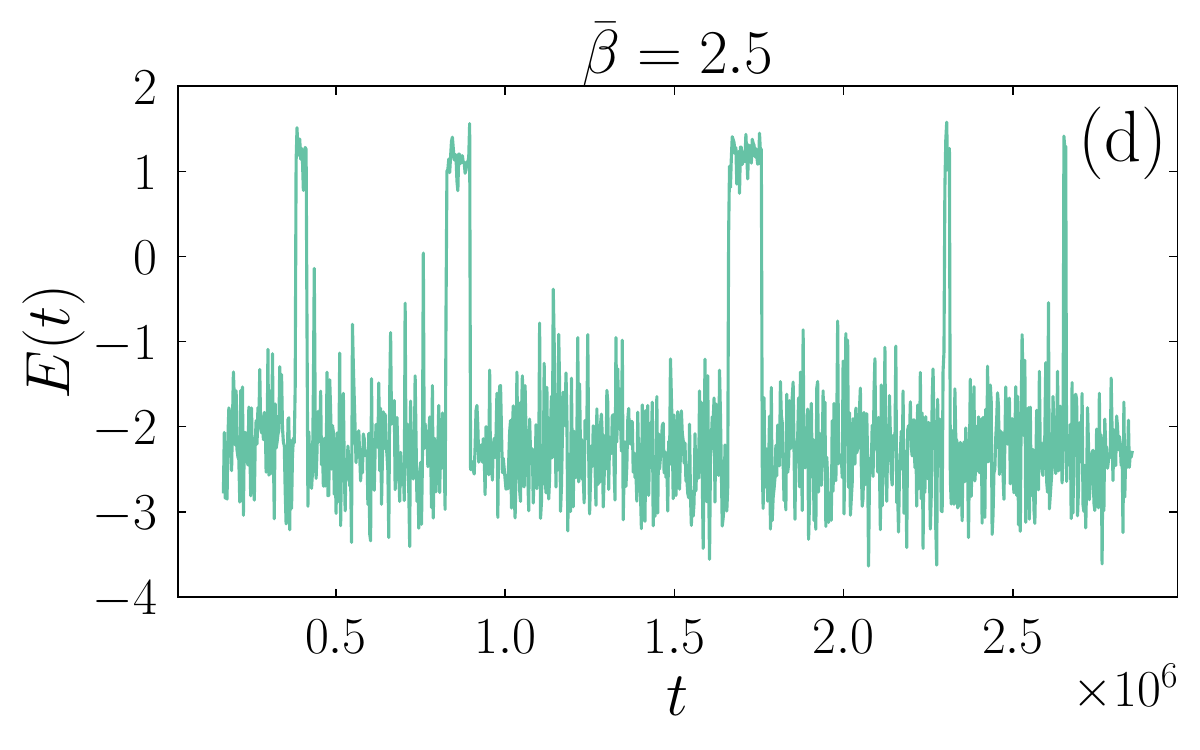}
}
\makebox[\linewidth]{%
  \includegraphics[width=8cm]{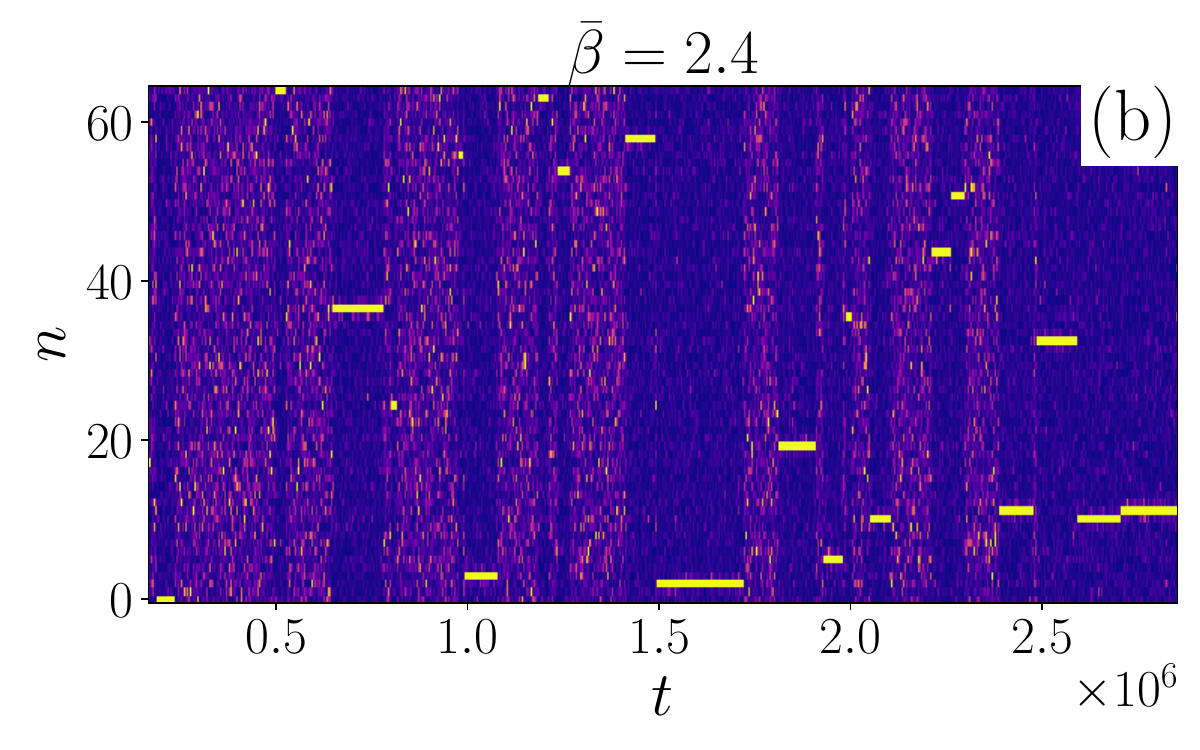} \hspace{0.1cm}
  \includegraphics[width=8cm]{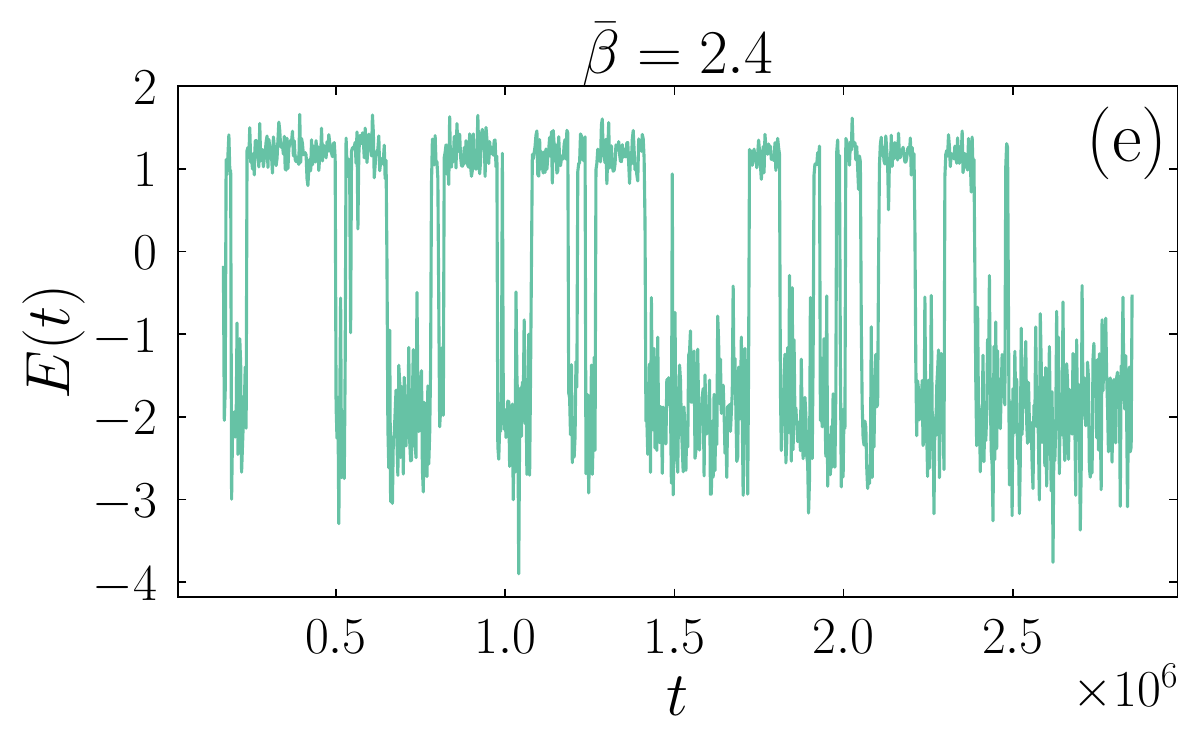}
}
\makebox[\linewidth]{%
  \includegraphics[width=8cm]{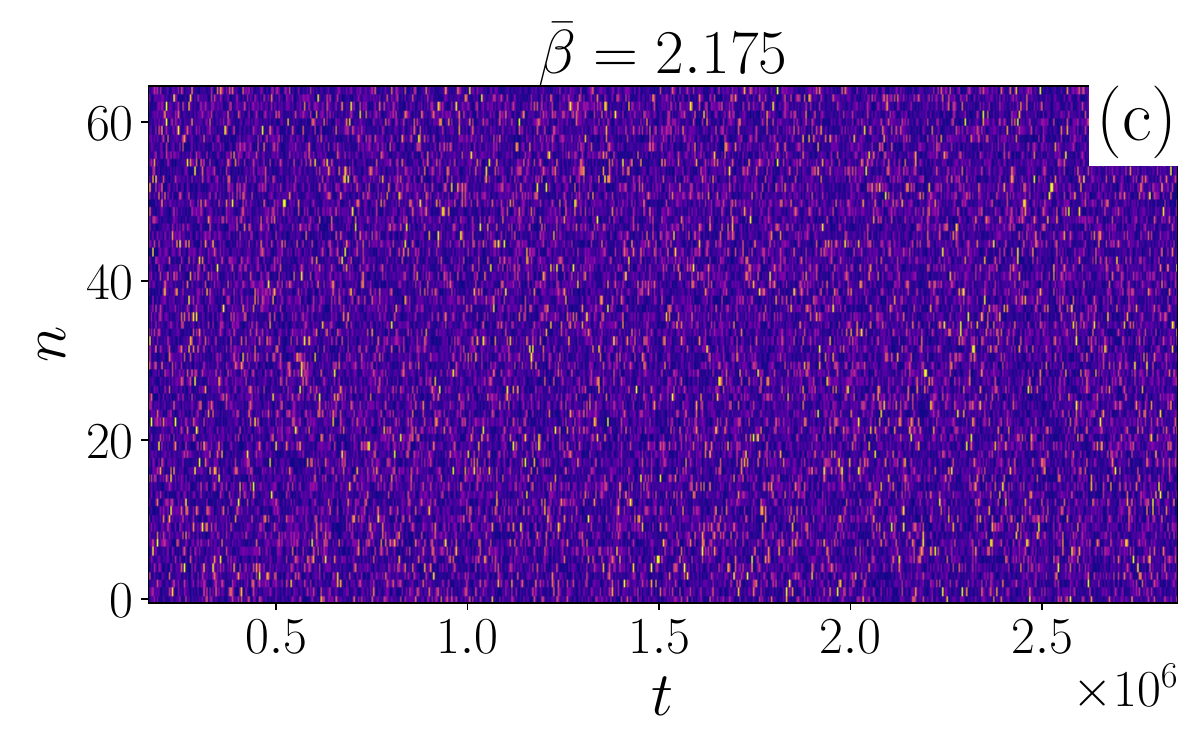} \hspace{0.1cm}
  \includegraphics[width=8cm]{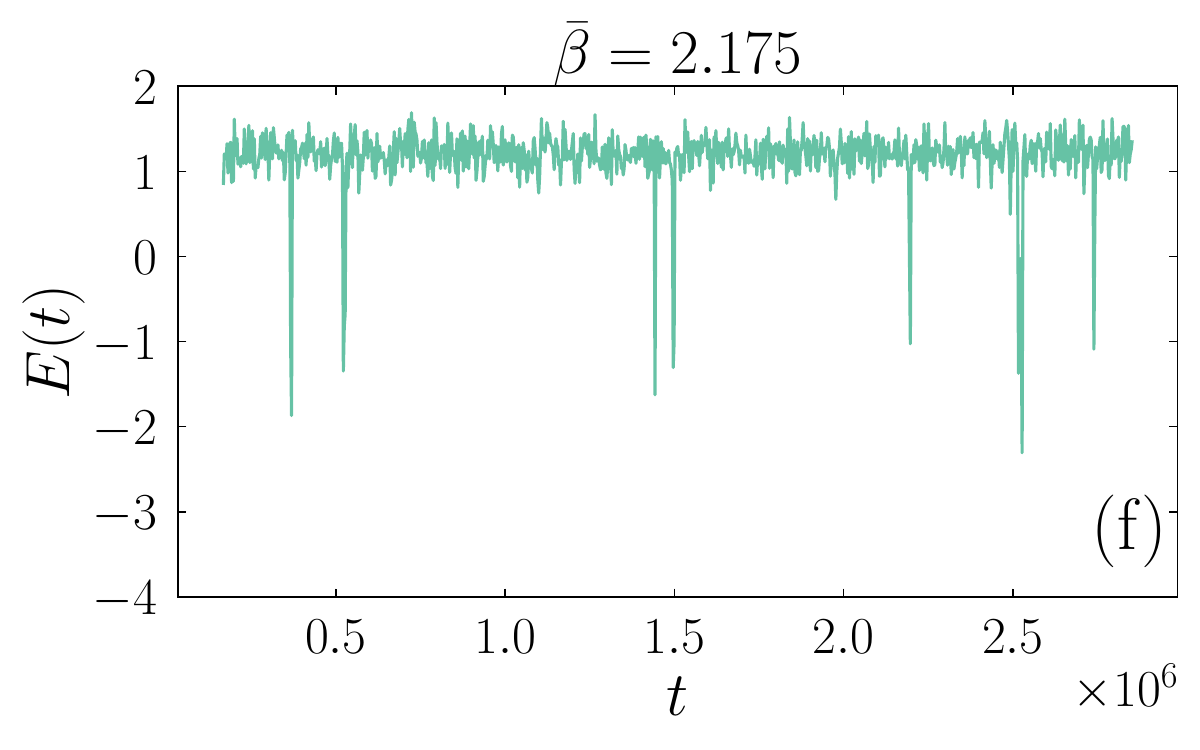}
}
\caption{
Space-time density plots of the pattern of weights, $|c_n(t)|^2$, 
in the stationary state as obtained from numerical simulations
of eq.~\eqref{2.2.1} with stepsize $\tau=0.01$ for a system of size $L=64$, noise strength $\sigma=0.3$ and $\alpha=16$. Scaled inverse temperature values:
(a) $\bar{\beta}_{low} = 2.175\quad(\beta_{low}=17.4)$, (b) $\bar{\beta}_{crit} = 2.4 \quad(\beta_{crit}=19.2)$, and
(c) $\bar{\beta}_{high} = 2.5 \quad(\beta_{high}=20)$(see as well Figure \ref{fig33}(a)), Colour coding:
from blue ($|c|^2=0.01$) to yellow ($|c|^2=0.99$). Panels (d), (e), and (f) show the corresponding time evolution of the total energy $E(t)$. Data have been obtained using a localised initial condition $c_n(0)=\delta_{n,32}$, skipping a transient of length about $\Delta t=1.6 \times 10^{5}$ to ensure stationary behaviour.}
\label{fig33}
\end{figure*}

In the following, we explore the phase transition and the nature of the phases in more detail. We will show results for 
systems of size $L=64$ with $\alpha=16$. To a large degree, the findings do not depend on these particular values, apart from time scales becoming excessive if the values are increased substantially. As shown in Figure \ref{fig32}(a) (full symbols, orange), the phase transition region extends between $\bar{\beta}_{low}=2.175$
($\beta_{low}=17.4$) and $\bar{\beta}_{high}=2.5$ ($\beta_{high}=20)$ with a maximum  variance at $\bar{\beta}_{crit}=2.4$
($\beta_{crit}=19.2$), see the black points in Figure \ref{fig32}(a). We will mostly use  these temperatures to illustrate signatures of  the low- and the high-$\beta$ phase, as well as the  critical behaviour in between. Figure \ref{fig33} shows 
space-time density plots of the time evolution of $|c_n(t)|^2$ in the stationary state, as well
as the corresponding time evolution of the energy $E(t)$. The low-$\beta$ phase 
with predominantly large values of energy is dominated by
a random pattern with predominantly small fluctuations of $|c_n|$, interspersed with short-lived localised peaks. The latter structures diminish if $\beta$ is lowered further.
The high-$\beta$ phase with predominantly low values of the energy
is dominated by localised structures which display a slow movement,
interrupted occasionally by low amplitude disordered phases. The movement of the peaks freezes when
$\beta$ is increased.  In the center of the phase transition region, one observes an intermittent switching
between these two phases, which is accompanied by large fluctuations of the energy.

\begin{figure}[h!]
\centering
\makebox[\linewidth]{%
  \includegraphics[width=8cm]{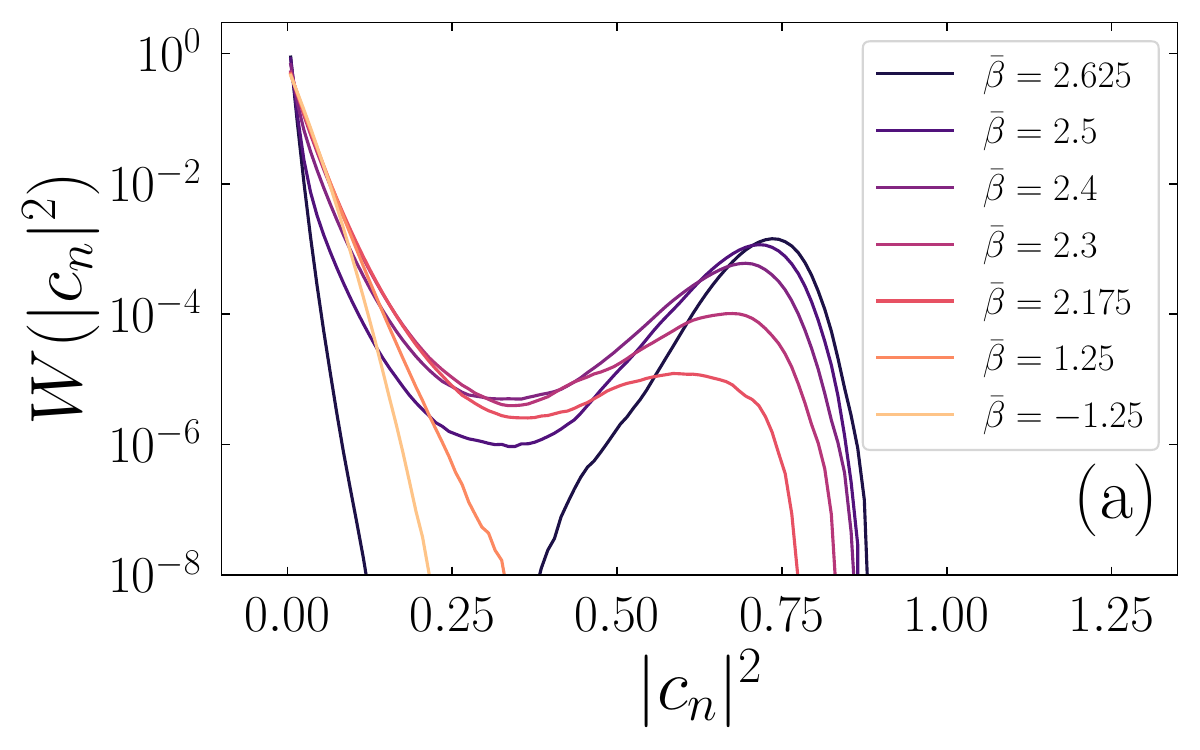} \hspace{0.1cm}
  \includegraphics[width=8cm]{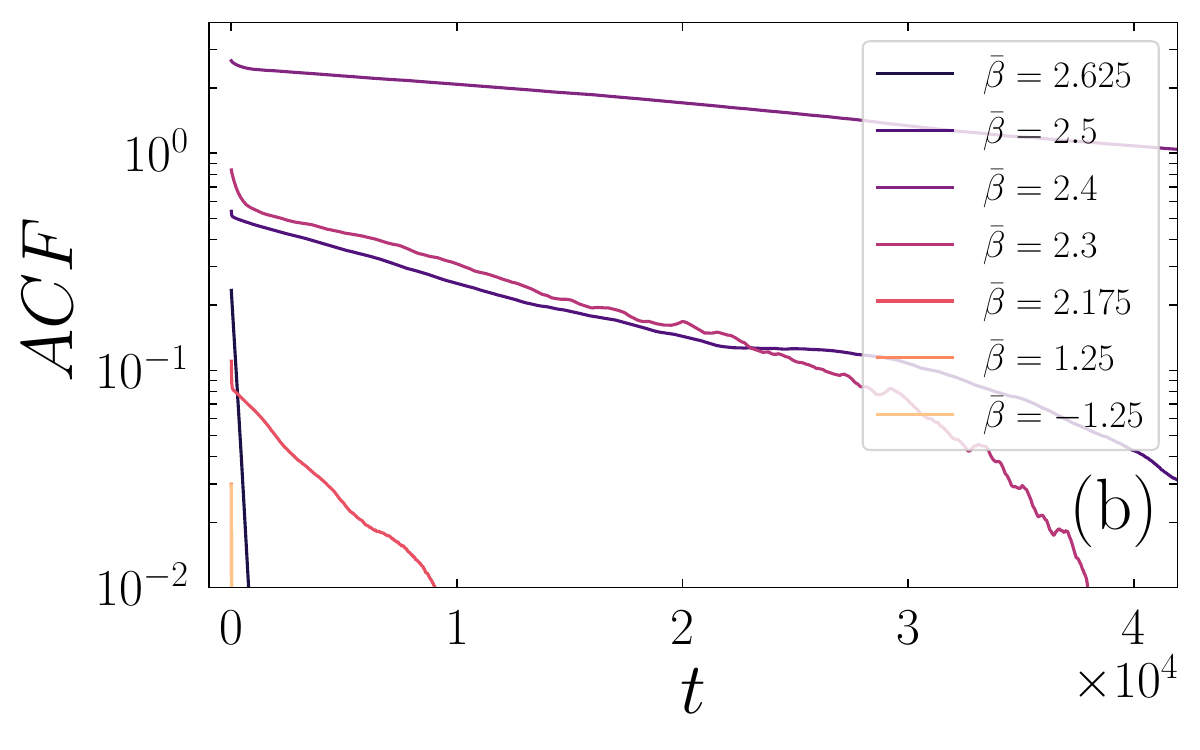}
}
\caption{
(a) Probability distribution $W(|c|^2)$ of weights in the stationary state, obtained
for a system of size $L=64$, noise strength $\sigma=0.3$, $\alpha=16$, 
and different values of the temperature, including the three values used in Figure \ref{fig33}(a-c). 
Data have been
computed from a time series of length about $T_{sim}=2.6 \times 10^{6}$ with sampling rate $\tau=0.01$, skipping a
transient of duration about $\Delta t=1.6 \times 10^{5}$ to ensure stationary behaviour. (b) Autocorrelation
function of the energy for the parameter setting of part (a), 
showing that the correlation time is longest in the middle of the phase transition region. 
Data have been
obtained from time series of length about $T_{sim}=2.6 \times 10^{6}$ skipping transients of length about $\Delta t=1.6 \times 10^{5}$.
}\label{fig34}
\end{figure}

For a quantitative characterisation of patterns and phases, we resort to a computation
of the distribution of weights, $W(|c|^2)$. The data in Figure \ref{fig34}(a)
show an exponentially decaying distribution for negative and sufficiently small positive $\bar\beta$ values, with a peak at large values of $|c|^2$ building up as the critical region is approached and becoming dominant
in the high-$\beta$ phase. These results of the stochastic model are completely in line with results reported for the Hamiltonian dynamics of the DNSE early on in
the context of negative temperature states
\cite{rasmussen_statistical_2000, rumpf_transition_2008}. Thus, the features of the stochastic 
motion cover important aspects of the deterministic model. A basic characterisation
of the temporal properties of the stationary behaviour is provided through
the energy autocorrelation function. In the low- and the high-$\beta$ phase energy 
correlations decay exponentially on a fast time scale (see Figure \ref{fig34}(b)), while, as expected, correlations slow down considerably in the critical region (cf. as well
the time traces in Figure \ref{fig33}, and Figure \ref{fig22}(b) for the dependence
of the correlation function on the noise strength $\sigma$ in the disordered phase).

\begin{figure}[h!]
\centering
\makebox[\linewidth]{%
  \includegraphics[width=8cm]{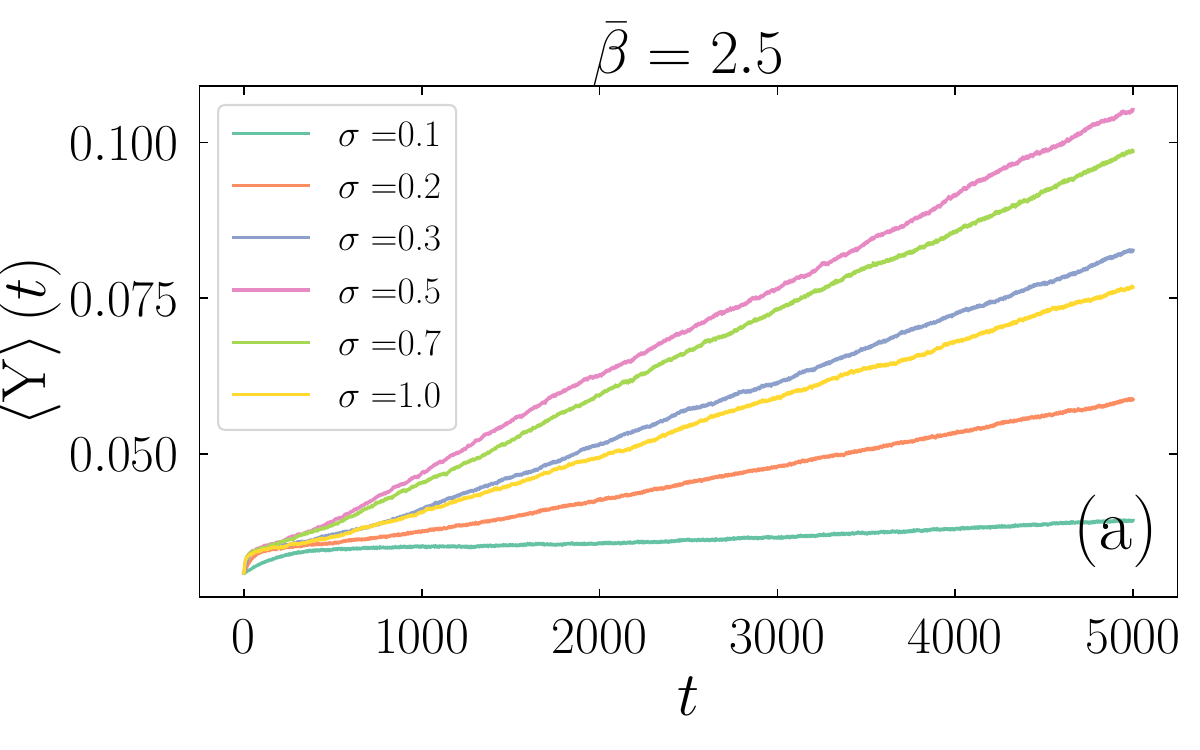} \hspace{0.1cm}
  \includegraphics[width=8cm]{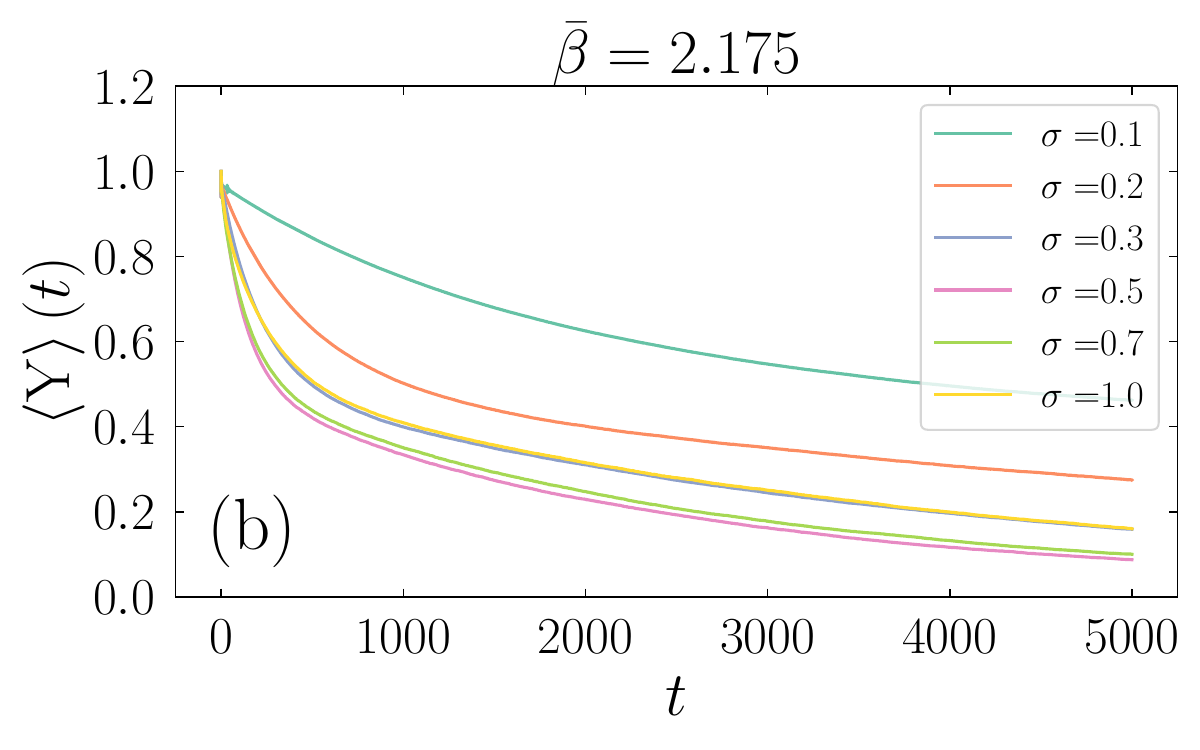}
}
\caption{
Transient time evolution of the mean inverse participation ratio 
$\langle Y\rangle (t)=\sum_{n=1}^L \langle |c_n(t)|^4 \rangle$
for a system of size $L=64$, $\alpha=16$ and different noise strengths:
$\sigma=0.1$ (cyan), $\sigma=0.2$ (orange), $\sigma=0.3$ (blue), $\sigma=0.5$ (pink), 
$\sigma=0.7$ (green), and $\sigma=1.0$ (yellow). The time evolution has been computed for an ensemble
of $10^4$ realisations of the stochastic process.
(a) Time evolution for $\bar{\beta}_{high} = 2.5\quad(\beta=20)$ (localised phase) with random initial conditions drawn
from the distribution $\delta(N-1)$. (b) Time evolution for $\bar{\beta}_{low} = 2.175\quad(\beta=17.4)$ (disordered phase)
with a localised initial condition $c_n(0)=\delta_{n,32}$.
}
\label{fig35}
\end{figure}

So far, we have focused on the stationary behaviour of the different phases. A key aspect of nonequilibrium dynamics consists in the formation of localised
structures in the localised phase starting from a random initial condition, as well as
the decay of localised structures in the disordered phase when starting from a localised initial
condition (cf. e.g. \cite{Bray_AP93} for a comprehensive overview of these aspects in the context of condensed matter physics).
A simple quantity that captures such an 
aspect is the so-called
inverse participation ratio $Y(t)=\sum_{n=1}^L |c_n(t)|^4$. Small values indicate
a delocalised structure, while the emergence of a localised structure is signalled by
inverse participation ratios approaching the maximum value of one. Figure \ref{fig35}
shows the time evolution of the inverse participation ratio in the two scenarios
mentioned above. If we start with high $\beta$ values, that means in the localised phase, 
from random initial conditions drawn from the distribution $\delta(N-1)$, the inverse
participation ratio $Y(t)$ increases with time, indicating that localised structures appear when approaching
the stationary state, Figure \ref{fig35}(a). One observes a non-monotonic dependence of the process on the noise strength $\sigma$, and there seems to appear an optimal noise strength which
supports the formation of localised structures. At first sight, such a kind of phenomenon reminiscent of 
noise-induced formation of structures \cite{HoLe:84} seems plausible.
However, looking at the low-$\beta$ case, i.e.~the random phase,
and starting from a localised initial condition $c_n(0)=\delta_{n,32}$ we see the decay
of this structure, signalled by a decay of the inverse participation ratio, see
Figure \ref{fig35}(b). Again, it seems that the same optimal noise strength speeds up this process. Unlike the simple diffusive picture mentioned in section \ref{sec:2.2} the impact of noise is more sophisticated in this setup and comes from an interaction of the
deterministic nonlinear dynamics with the stochastic force in eq.~\eqref{2.2.1}, which so
far defies a simple heuristic reasoning, let alone a proper formal explanation.

\section{Discussion and Conclusion} \label{sec:4}
Our computer simulations for the DNSE coupled to a heat bath have shown that there is a phase transition in the positive-temperature regime from a disordered phase to a localised phase. 
The location of the phase
transition depends essentially on the parameter $\bar\beta = \alpha\beta/2L$. This phase transition can be
understood by the fact that a system coupled to a heat bath minimises its free energy $F=E-TS$. For high
temperatures, the disordered phase is preferred since the product $TS$ becomes large. In the 
low-temperature phase, $TS$ is small, and the phase that minimises energy is chosen. The phase transition
occurs when the energy difference $\Delta E$ between the two phases equals $T$ times  the entropy
difference $\Delta S$, giving $\alpha \propto T L$ because the energy scale is set by $\alpha$ and because
entropy is extensive, implying that $\bar\beta \propto \beta \alpha/L$ is the relevant control parameter,
in agreement with our computer simulations (see Figure \ref{fig32}). 

The gist of the phase transition itself can be captured by a simple mean-field model that gives an approximate expression
of the canonical partition function $Z$ in dependence on a suitable parameter. Minimization of
$F = -k_B T\ln Z$ with respect to this parameter then yields the thermodynamically stable phases. 
We assume that $\alpha$ is large enough so that in the low-temperature phase the breather can be approximated by a state where, on average, the weight $|c_n|^2$ on one site is larger than on the others.
We implement this by setting $\langle|c_1|^2\rangle = a$ and $\langle|c_n|^2\rangle = b$ for $n>1$, where
normalisation requires $\langle N \rangle=a + (L-1)b=1$. This ansatz accommodates both phases as the high-temperature phase is characterised by $a=b=1/L$. The low-temperature phase breaks translational symmetry since the mean weight at one site differs from that of the others. The difference $a-b$ can be interpreted as the order parameter associated with the phase transition. The normalisation condition constrains the parameter $b$ to the range $[0,1/L]$ to leading order
in system size, as $a\geq 0$ and $b\geq 0$.
The energy of the system can be estimated by 
\begin{equation}
E(b) = 2-\frac{\alpha}2(b^2(L-1)+(1-(L-1)b)^2)
\simeq  2-\frac {\alpha}2(1-x)^2
\label{4.1a}
\end{equation}
keeping only leading-order terms in $L$ and using the abbreviation $x=b L\in[0,1]$. In writing down
eq.~\eqref{4.1a} we neglected the nearest-neighbour coupling terms. 
To obtain the partition integral $Z$, we estimate the degeneracy of each energy value by making the plausible assumption that the range of the weights $|c_n|^2$ is of the same order as their mean, which is $b$.
We thus obtain the estimate
\begin{equation}
Z \simeq b^L e^{-\beta E(b)} \simeq \frac{x^L}{L^L} e^{(\beta \alpha/2)(x-1)^2} e^{-2 \beta}
\end{equation}
resulting in 
\begin{equation}\label{4.4a}
\frac{\beta F}{L}=-\frac{1}{L}\ln Z =\Phi(x)+\ln(L)+2\beta/L
\end{equation}
where 
\begin{equation}
\label{4.4b}
\Phi(x)=-\ln x -\bar{\beta}(x-1)^2
\end{equation}    
abbreviates the nontrivial part of the free energy, since the other terms in eq.~\eqref{4.4a}
drop out when the minimum of $F$ is calculated among the accessible values $x\in[0,1]$.

\begin{figure}[h]
\centering
\makebox[\linewidth]{%
  \includegraphics[width=8cm]{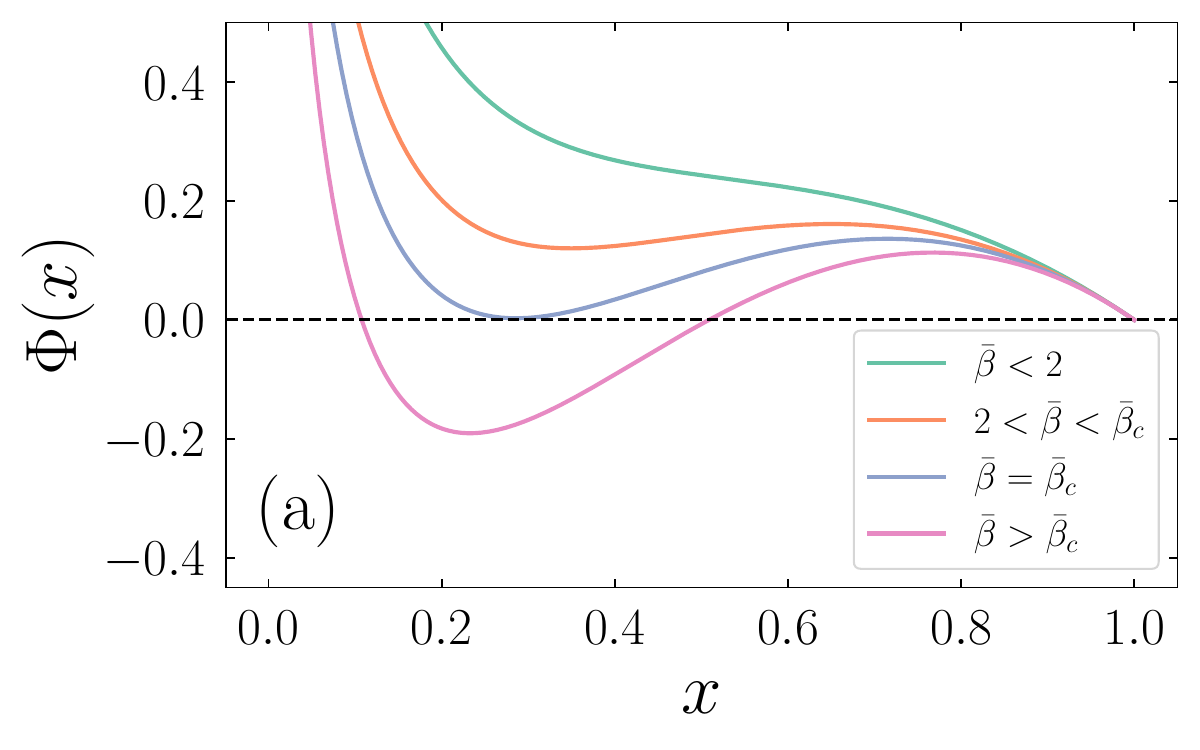} \hspace{0.1cm}
  \includegraphics[width=8cm]{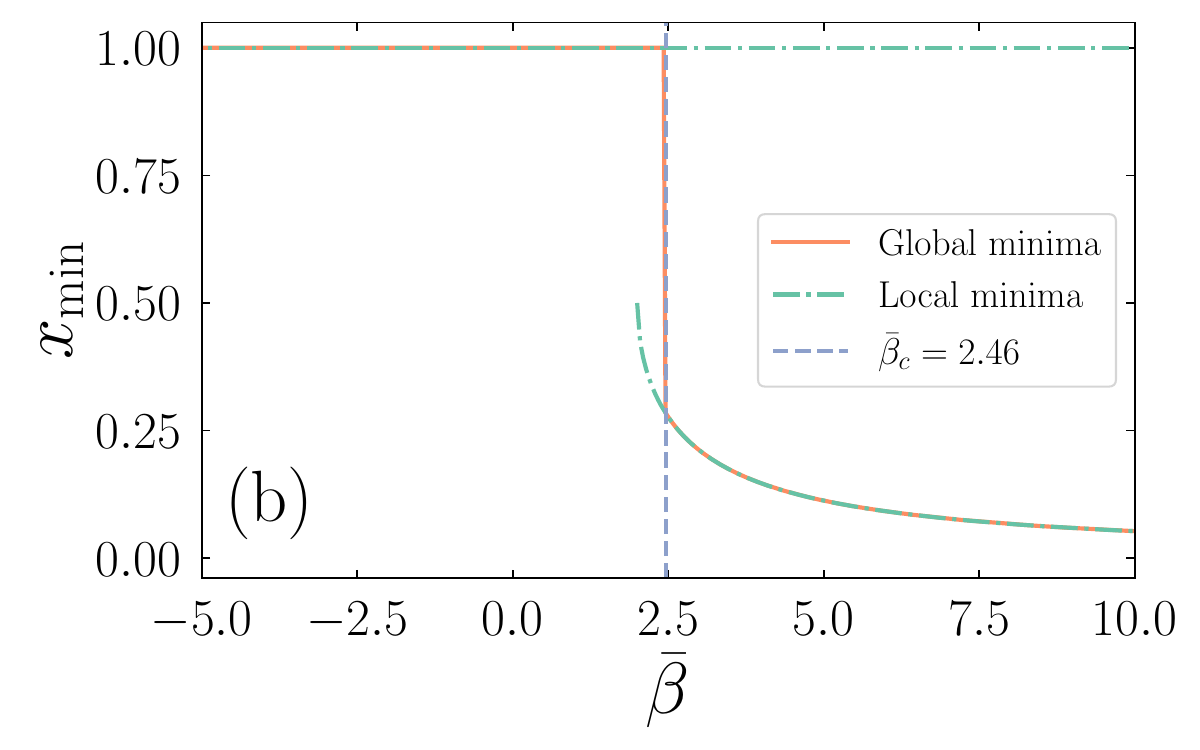}
}
\caption{
(a) Dependence of the nontrivial part of the free energy, eq.~\eqref{4.4b}, on the order parameter $x$, for
different values of the scaled inverse temperature $\bar{\beta}$.
(b) Global (solid) and local (dash-dotted) minima $x_{min}$ of the free energy, eq.~\eqref{4.4b},
in dependence of the scaled inverse temperature $\bar{\beta}$. The dashed vertical line 
indicates the critical value $\bar{\beta}_c$.
}\label{fig36a}
\end{figure}

The scaled free energy, eq.~\eqref{4.4b} has a global minimum at $x=1$ for $\bar{\beta}<2$
and develops a second local minimum at $\bar{\beta}=2$ which becomes the global minimum for
$\bar{\beta}> \bar{\beta}_c \simeq 2.46$, see Figure \ref{fig36a}. The
value of $x= Lb$ that minimises $F$ jumps from $1$ to the vicinity of $0.3$ at the critical
temperature determined by $\bar\beta_c$. This means that approximately 70 percent of all weight is concentrated in one site. The jump of $x$ (and therefore the jump of the order parameter $a-b$) means that the phase transition is of first order. The inner minimum of the free energy \eqref{4.4b} becomes metastable for $2<\bar{\beta}<\bar\beta_c$ but persists down to $\bar{\beta}=2$, where this minimum vanishes. The outer minimum at 1 remains metastable for $\beta>\bar{\beta}_c$.
The energy associated with the global minimum $x_{min}$ is obtained from eq.~\eqref{4.1a} and shown above in Figure \ref{fig32}(b) as a black dash-dotted line. The location $\bar\beta_c$ of the jump agrees pretty well with the simulation data in this figure, and so does the entire low-temperature phase, 
$\bar{\beta}>\bar{\beta}_c$, including the metastable region. 

The energy of the high-temperature phase, $\bar{\beta}<\bar{\beta}_c$, resulting from our 
mean-field theory remains at the infinite-temperature value 2 when $\bar{\beta}$ increases, 
see Figure \ref{fig32}(b), because the theory neglects the coupling
between neighbours, in contrast to the simulation data, which show a decreasing energy. 
However, there is a good match between the high-temperature energy and the analytical result obtained for the case $\alpha=0$, which
neglects the nonlinearity but takes into account the coupling between neighbours. The resulting expression for the energy (eq.~\eqref{e.16} in appendix \ref{sec:e})  agrees well with the simulation data shown in Figure \ref{fig31}(a) (black dashed line), indicating that the nonlinear term 
proportional to $\alpha$ has a very small influence on the high-temperature phase. 
This is plausible since in the disordered phase the contribution of the nonlinear term to the energy is of order $\mathcal{O}(L^{-1})$ as compared to the normalisation.

Interestingly, our results for the order parameter $x_{min}$, for the critical inverse scaled temperature $\bar\beta_c$, and for the free energy function $\Phi(x)$ evaluated at $x_{min}$ agree (apart from an additive constant in the free energy) with the results of an exact analytical calculation for the full DNSE (including nearest-neighbor coupling) performed by Chatterjee and Kirkpatrick \cite{chatterjee2012probabilistic} in the limit where the number of lattice sites goes to infinity at fixed average weight per site (corresponding in our notation to a fixed $\alpha/L$). This means that our simple mean-field theory becomes exact in this limit. 

Due to finite-size effects, our simulations do not show a jump at the phase
transition but a continuous decline that becomes steeper with increasing
system size (cf. Figures \ref{fig31}(a) or \ref{fig32}(b)). Energy values
between
those of the high- and low-temperature phases are realised by
a "phase coexistence", with dynamics switching between time intervals
that show the disordered phase and those that show the localised phase,
see Figure \ref{fig33}. This coexistence is only visible when the system size is small enough that over the time periods accessible to simulations, several switches between the two phases occur. Otherwise, we have a nonequilibrium situation where the system remains metastably in one of
the phases, leading to the hysteresis seen in Figure \ref{fig32}(b)
for $L=128$. 

The instability of the homogeneous state of the system over a large energy interval might help to understand
results obtained in microcanonical simulations, where the energy is constrained to have the same value at all times. According to our
mean-field theory, such
energy values can be realised by an $x=bL$ value that does not
minimise the free energy $F$. This state is already a symmetry-broken
state since the mean weight at one site is larger than that of the other sites. In the canonical
ensemble, this solution becomes unstable and is replaced by a phase
coexistence, a situation that bears some resemblance to the Maxwell
construction performed for the solution of the van der Waals equation
in the parameter range where the homogeneous solution is unstable.

The mean field approach captures the phase transition quite well, even
at the quantitative level. The phase diagram of the DNSE is normally
given in terms of the energy and another system parameter, see for instance
Figure \ref{figb1}, since the analysis is usually based on
a Hamiltonian, i.e., a microcanonical approach. To translate our result to
such a phase plane, the caloric equation of state is needed. Fortunately, our
results give sufficient information to perform such a step.
By combining the mean-field theory with the
caloric equation of state for $\alpha=0$ we can obtain the line that
bounds the localised phase in the $E$-$\alpha$ phase plane.
The mean-field theory tells us that the metastable localised
low-temperature (i.e., high $\bar\beta$) phase ceases
to exist at $\bar{\beta}= \alpha\beta/2L=2$.
If we evaluate eq.~\eqref{e.15}, which applies to the high-temperature phase, with $\beta=4L/\alpha$ we obtain the critical
energy
\begin{equation}\label{4.5a}
E_c(\alpha)=2-\frac{16}{\alpha + \sqrt{\alpha^2+64}}, \quad (\alpha>0) \, .    
\end{equation}
For energies $E>E_c(\alpha)$ the DNSE is in the disordered phase whereas for
$E<E_c(\alpha)$, the DNSE is either in the phase coexistence region or in the localised phase.
This critical line $E_c(\alpha)$ is shown in the phase diagram
Figure \ref{figb1} together with its conjugate counterpart for
$\alpha<0$. This theoretical prediction agrees quite well
with results obtained from numerical simulations.
The data points next to this line have been obtained from
simulations in a system of size $L=1024$ with noise strength
$\sigma=0.3$ skipping transients of duration $\Delta t=3.4 \times 10^5$
and computing energy averages over a time interval of length
$T_{sim}=3.4 \times 10^5$ (to cope with stiffness due to large $\beta$ values
the midpoint step in the numerical integration procedure, see
appendix \ref{sec:d}, has been
occasionally replaced by more stable schemes).
The dots indicate the energy the system jumps to
when the low-temperature phase in numerical simulations becomes unstable
in an adiabatic parameter downsweep.

A phase transition between a localised breather phase and a disordered phase has also been
reported previously in several publications that are based on analytical calculations and computer
simulations of the microcanonical ensemble in the limit where the nearest-neighbour
interaction is
neglected \cite{GrIuLiMa_JSM21,gradenigo2021condensation,giachello2025localization}.
These publications use a dimensionless version of the model eq.\eqref{2.1} where the
nonlinear term has a positive sign, and the breather state therefore occurs for negative
temperatures. In the limit of infinite system size, the authors find a phase transition
line that coincides with the infinite-temperature line, with finite-size corrections
of order $\mathcal{O}(L^{-1/3})$
moving the transition line into the negative-temperature region, see for instance
Figure 1 in \cite{gradenigo2021condensation}.
In fact, an analogous phenomenon happens
in our approach as well, since a thermodynamic limit along the lines of
\cite{GrIuLiMa_JSM21,gradenigo2021condensation,giachello2025localization} requires
$\alpha$ to be of the order $\mathcal{O}(L)$, see for instance \cite{EBRAHIMI2025134905}, so that
again the infinite temperature line $E=2$ is approached when $\alpha$ becomes large, see eq.~\eqref{4.5a}.
Of course, any additional finite-size correction depends crucially on the details of the thermodynamic limit, so that we do not recover
any correction of order $\mathcal{O}(L^{-1/3})$. However, even though the phase transition line
approaches the infinite temperature line $E=2$ when $\alpha$ is proportional to $L$, our analysis shows that the phase transition
temperature given by $\beta_c = 2\bar\beta_c (L/\alpha)$ remains finite, due to the particular form of the caloric equation of
state (cf.~eq.~\eqref{e.16} for the simpler case without interaction potential).

Our stochastic equation of motion,  eq.~\eqref{2.2.1}, includes the noise strength $\sigma$ as an additional parameter. 
 The non-monotonic dependence of
the transient behaviour on
$\sigma$, see Figure \ref{fig35}, resulting in an optimal noise strength, is an
unexpected property. Such a feature might be caused by a nontrivial interaction
between the noise and internal time scales of the deterministic
dynamics, reminiscent of stochastic resonance. 
Stochastic resonance is normally triggered by a resonance phenomenon
between two time scales, often caused by a deterministic external drive and
an internal stochastic motion which depends on noise strength,
resulting in an enhanced periodic signature in output signals
occurring at on optimal noise strength.
In our case we observe a non-monotonic dependence of transient times
on the noise strength. Transient times become minimal at
an optimal noise strength, a feature which is certainly linked
with the internal dynamics of the DNSE.
Having said that there is no obvious resonance phenomenon between
two different time scales visible.
A proper explanation, which is beyond the scope of the present paper, may
shed some light on the intricate nonequilibrium properties of the DNSE.

The phase transition of the DNSE is traditionally related to negative temperatures,
as the vast majority of investigations use a normalised version of the model with
a repulsive on-site potential. Our computations show that such phase transitions
persist if the DNSE with attractive on-site potential is coupled to a heat bath,
since the sign of the potential, i.e., switching between the focusing and
the defocusing version of the DNSE, can be used to change effectively the sign of the
temperature. As already mentioned in the literature, the additional constant of motion
is crucial for the phase transition to occur in this one-dimensional model.
Otherwise, the usual transfer operator approach yields compact operators and
analyticity of the free energy in the thermodynamic limit. The current investigation
shows that one should expect the phase transition to occur at finite temperatures,
and the model studied here may trigger further attempts to find such
phase transitions in real-world experiments.

\section*{Funding and Competing Interests}
 This work was supported by the German Research Foundation (DFG) under contract number Dr300/16 and via CRC 1270 “Electrically Active Implants”, Grant/Award Number SFB 1270/2-299150580.

\data{No new data were created or analysed in this study.}

\suppdata{None.}
 
\appendix

\section{Microscopic derivation}\label{sec:a}
For the convenience of the reader, we give here a brief account
how the stochastic differential equation, eq.~\eqref{2.2} can be derived from a 
microscopic model
when the Hamiltonian eq.~\eqref{2.1} is coupled to a heat bath.
The full Hamiltonian $H$ then includes three contributions: the Hamiltonian 
$H_S$ of the subsystem, the 
Hamiltonian $H_B$ of the heat bath, and the coupling between subsystem and bath $H_{SB}$, 
$H = H_S + H_B + H_{SB}$. By eliminating the bath degrees of freedom from the 
equation of motion, we will derive the effective equation of motion of the subsystem, 
which includes a stochastic noise term and a damping term. The elimination
process follows a well established procedure using projection operator techniques developed
more than half a century ago, \cite{Naka_PTP58, Zwan_JCP60}. Overviews of this
approach can nowadays also be
found in standard textbooks, e.g. \cite{FiSa:90}. One does not even have to specify the
details of the heat bath, as long as some basic properties are satisfied. However,
excessive formalism can be avoided and one may even be able to provide a
rigorous account, see for instance \cite{FoKaMa_JMP65}, if one constrains to a special 
heat bath consisting of harmonic oscillators
\begin{equation}
    \label{a2.2.2}
    H_{B} = \sum_q \omega_q \bar{b}_q b_q  \, ,
\end{equation}
where $q$ denotes the wavenumers,  $b_q$ the
phonon amplitudes, and $\omega_q$ the dispersion relation. Overall, the dynamics
should still preserve the normalisation $N$, which puts a constraint on the coupling
between the subsystem and the heat bath. Here we chose the lowest-order term that is compatible 
with the symmetries and conservation laws of the system, which gives a coupling that 
is linear in the phonon amplitudes $b_q$ and linear in the weights $|c_n|^2$ (to preserve the normalisation $N$),
 \begin{eqnarray}
     \label{a2.2.3}
     H_{SB} &=& \sum_{n=1}^L |c_n|^2 (B_n + \bar B_n) \\
     B_n &=& \sum_q g_q e^{-inq} b_q \, .
\end{eqnarray}
The coupling constants $g_q$ contain the interaction details, and the complex exponential ensures a (lattice) translation-invariant Hamiltonian. In fact, such a kind of coupling
mimics the electron-phonon interaction in solids. If needed, more realistic Hamiltonians
can be adopted as well.
The time evolution of the variables $b_q$ and $c_n$ is calculated using the Poisson bracket
\begin{eqnarray}
\label{a.1}
\{F, G\} &=& \sum_n i\Big( \frac{\partial F}{\partial \bar{c}_n}
\frac{\partial G}{\partial c_n} - \frac{\partial F}{\partial c_n}
\frac{\partial G}{\partial \bar{c}_n}\Big) + \sum_q i \Big(\frac{\partial F}{\partial \bar{b}_q}
\frac{\partial G}{\partial b_q} - \frac{\partial F}{\partial b_q}
\frac{\partial G}{\partial \bar{b}_q} \Big)\, .
\end{eqnarray}

The time evolution of the phonon amplitudes is governed by a linear inhomogeneous
differential equation $\dot{b}_q = \{H, b_q\}=i \partial H/\partial \bar{b}_q$ which can be formally integrated
\begin{align}
\begin{split}
\label{a2.2.4}
   b_q(t) &= e^{i\omega_q t} b_q(0) + i \int_0^t e^{i \omega_q (t-s)}\sum_m \bar{g}_q e^{imq}|c_m(s)|^2 \dd{s} \, .\\
\end{split}
\end{align}
This result can be plugged into the equation of motion $\dot{c}_n=\{H,c_n\}=i \partial H/\partial \bar{c}_n$
in order to eliminate the variables $b_q$. We obtain in addition to the DNSE time evolution, a fluctuating residual force caused by the initial conditions $b_q(0)$ and a memory
term containing a convolution integral, both of which are due to the coupling to the heat 
bath
\begin{eqnarray}
\label{a2.2.5}
    \dot{c}_n(t)
    &=& i \frac{\partial H_S}{\partial \bar{c}_n}+ i c_n(t) f_n(t) + i c_n(t) \sum_m \int_0^t \gamma_{n-m}(t-s) |c_m(s)|^2 \dd{s}\, . 
\end{eqnarray}
Here, the real-valued residual force is given by
\begin{equation}
\label{a2.2.6}
    f_n(t) = \sum_q \left(g_q e^{-inq} e^{i\omega_q t} b_q(0) + 
    \bar{g}_q e^{inq} e^{-i\omega_q t} \bar{b}_q(0)\right)    
\end{equation}
while the real-valued memory kernel reads
\begin{align}\label{a2.2.7}
\gamma_{n-m}(t-s)
= i \sum_q |g_q|^2 \Big[
    e^{-i(n-m)q} e^{i\omega_q (t-s)}
    - e^{\,i(n-m)q} e^{-i\omega_q (t-s)}
\Big].
\end{align}
We assume that the heat bath is in thermodynamic equilibrium with the inverse temperature 
$\beta$ so that the distribution of the phonon amplitudes is given by $\exp(-\beta H_B )/Z_B$. That means the initial conditions $b_q(0)$ are independent Gaussian random variables with
\begin{equation}
\label{a2.2.8}
    \left\langle \bar{b}_q(0) b_{q'}(0) \right\rangle = \delta_{qq'}\frac{1}{\beta\omega_q}, \quad \left\langle b_q(0) b_{q'}(0) \right\rangle = 0 \, .
\end{equation}
Therefore, the residual forces $f_n$ are Gaussian variables as well, and their correlation 
function is given by

\begin{align}
\label{a2.2.9}
\begin{split}
    \left\langle f_n(t) f_m(s) \right\rangle &= \sum_q \frac{|g_q|^2}{\beta \omega_q} \Big(e^{-i(n-m)q} e^{i \omega_q (t-s)} + e^{i(n-m)q} e^{-i \omega_q (t-s)} \Big) = \Gamma_{n-m} (t-s) \, .
\end{split}
\end{align}

We will adopt later the usual assumption that the correlations in the heat bath decay fast due to the very large number of bath degrees of freedom, so that the expression \eqref{a2.2.9} practically vanishes after a short bath correlation time $\tau_c$, and the residual forces can be effectively viewed as white noise. 

Comparing the expression \eqref{a2.2.9} for the noise correlation function with expression \eqref{a2.2.7} for the memory kernel, we obtain the relation
\begin{equation}
\label{a2.2.10}
    \gamma_{n-m}(t-s) = \beta \frac{d \Gamma_{n-m} (t-s)}{dt} \, .
\end{equation}    
This equation constitutes a fluctuation-dissipation theorem (or a generalised Einstein relation) for our model. It ensures that noise and damping are consistent with thermodynamic equilibrium. Substituting relation \eqref{a2.2.10} in the right hand side of eq.~\eqref{a2.2.5} gives, using an integration by parts and considering times large compared to $\tau_c$

\begin{align}
    \label{a2.2.11}
    \dot{c}_n(t)
    &= i \frac{\partial H_S}{\partial \bar{c}_n}  + i c_n(t) f_n(t)\nonumber -i\beta c_n(t)\sum_m \Gamma_{n-m} (0)|c_m(t)|^2\\
    &+ i\beta c_n(t) \sum_m \int_0^t \Gamma_{n-m} (s)(\dot{c}_m(t-s)\bar{c}_m(t-s) + c_m(t-s)\dot{\bar{c}}_m(t-s)) ds \, .
\end{align}

The term containing $\Gamma_{n-m}(0)$ represents a bath-induced interaction that takes the 
(non-dissipative) impact of the heat bath in thermodynamic equilibrium into account. This term can in fact be absorbed in the first term of eq.~\eqref{a2.2.11} by making the replacement 
\begin{equation}
\label{a2.2.13}
H_S \to     H_S^{\textit{eff}} = H_S -\frac{\beta}{2}\sum_{n,m} \Gamma_{n-m}(0) |c_n(t)|^2|c_m(t)|^2 \, .
\end{equation}
This effective Hamiltonian can also be obtained by calculating the reduced density of the canonical equilibrium
\begin{equation}
\label{a2.2.12}
e^{-\beta H_S^{\textit{eff}}} / Z_{\textit{eff}} = 
\int   e^{-\beta H}/Z \prod_q d(b_q,\bar{b}_q)    \, 
\end{equation}
by making the substitution 
$b'_q = b_q + \bar{g}_q/\omega_q \sum_n \exp(iqn)(c_n + \bar{c}_n)$
to evaluate the integration over the heat bath variables.
The effective Hamiltonian captures all the equilibrium properties of the system
including, among others, a bath-induced effective interaction in the subsystem.
Here we assume, for simplicity,
that the coefficients $\Gamma_{n-m}(0)$ are either zero or small, so that
the contribution to the effective Hamiltonian can be neglected.

For the integro-differential equation \eqref{a2.2.11} we now require, as usual, a time
scale separation between the heat bath correlation time $\tau_c$
and the time scales of the subsystem, so that a so-called Markov
approximation applies. Because of the fast decay of the bath correlation $\Gamma_{n-m}(s)$ we
can then replace $c_m(t-s)$ by the instantaneous value $c_m(t)$ and extend the range of
integration to infinity. Formally, that means we replace the bath correlation functions
by $\delta$-functions. Then eq.~\eqref{a2.2.11} results in
\begin{eqnarray}
    \label{a2.2.11a}
    \dot{c}_n(t)
    &=& i \frac{\partial H_S}{\partial \bar{c}_n}  + i c_n(t) f_n(t)
     + i \beta \frac{\sigma^2}{2}  c_n(t) \left(\dot{c}_n(t)\bar{c}_n(t) + c_n(t)
    \dot{\bar{c}}_n(t)\right)  
\end{eqnarray}
with
\begin{equation}
\label{a2.2.17}
    \left\langle f_n(t) f_m(s) \right\rangle = \sigma^2 \delta_{n,m}\delta(t-s)
\end{equation}
and
\begin{equation}
\label{a2.2.18}
    \sigma^2 = \int_{-\infty}^{\infty} \Gamma_0(t) dt \, .
\end{equation}
Here, we have assumed, for simplicity, that any bath-induced interaction terms caused
by $\Gamma_{n-m}(s)$ with $n \neq m$ can be neglected, which is an implicit assumption about the coupling coefficients $g_q$. The last term in eq.~\eqref{a2.2.11a}
which represents the
dissipation in the system is a priori a Gilbert-type damping term (see e.g. 
\cite{Kiku_JAP56} and \cite{ElFi_PA83},
or \cite{HiMO_PRL09}
for a recent
discussion of the issue). We may solve eq.~\eqref{a2.2.11a} for the time derivatives
if we cast that equation in the form
\begin{eqnarray}\label{a2.2.19}
 \left(
\begin{array}{cc}\displaystyle  1-i \beta \frac{\sigma^2}{2} |c_n(t)|^2 & -i \beta \frac{\sigma^2}{2}
c_n^2(t)\\
\displaystyle i \beta \frac{\sigma^2}{2} \bar{c}_n^2(t) & 1+i \beta \frac{\sigma^2}{2} |c_n(t)|^2
\end{array}
\right)
\left( \begin{array}{c} \dot{c}_n(t) \\ \dot{\bar{c}}_n(t) \end{array}\right)
=
\left(\begin{array}{c}
\displaystyle i \frac{\partial H_S}{\partial \bar{c}_n} + i\sigma c_n(t) \xi_n(t) \\
\displaystyle -i \frac{\partial H_S}{\partial c_n} - i \sigma \bar{c}_n(t) \xi_n(t)
\end{array}
\right)
\end{eqnarray}
where we have used as well the abbreviation $f_n(t)=\sigma \xi_n(t)$ for the residual force.
The matrix on the right-hand side has a determinant of one, so that we can easily solve
for $\dot{c}_n(t)$ which gives the result eq.~\eqref{2.2}.
The damping term of this equation satisfies a fluctuation-dissipation relation 
with the noise term, so that the canonical thermodynamic equilibrium determines 
the stationary behaviour of the model (see appendix \ref{sec:c}). It is also worth noting that for the case of a heat bath consisting of harmonic oscillators, we do not have to invoke the Born approximation, so that the
resulting SDNSE even holds in the strong coupling case.

\section{Phase diagram of the DNSE}\label{sec:b}

The phase diagram of the DNSE, i.e., the bounds for the energy and a representation
of the different regimes in the parameter plane, is well established in the literature.
The considerations are often based on a standardised version of the DNSE
(see, for instance, \cite{ rasmussen_statistical_2000, rumpf_simple_2004, BaIuLiVu_PR21}) and those results can be easily translated to our setup
using suitable parameter transformations (see e.g.~\cite{EBRAHIMI2025134905} for the
transformation formulas). For the convenience of the reader, we here
summarise the elementary estimates within our notation of eq.~\eqref{2.1}. The two 
parameters in our setup are the energy $E=H_S$ and the on-site potential strength
$\alpha$. While the phase diagram summarises the properties of $H_S$ subjected
to the constraint $N=1$, the diagram also applies for the SDNSE, eq.~\eqref{2.2.1},
if one is able to translate between temperature and energy using the caloric equation of state,
since the stochastic dynamics respects the conservation law $N=1$ (see section \ref{sec:2.1}
and appendix \ref{sec:e}). 

Before establishing upper and lower bounds, we first express the Hamiltonian \eqref{2.1}
in a more convenient form. Using the identity
\begin{equation}\label{b.1}
\pm(\bar{c}_{k-1} c_k+ c_{k-1} \bar{c}_k) = |\bar{c}_{k-1}\pm c_k|^2 
- |c_k|^2-|c_{k-1}|^2
\end{equation}
together with $N=1$ we obtain, thanks to periodic boundary conditions,
for the Hamiltonian \eqref{2.1} the expressions
\begin{eqnarray}\label{b.2}
H_S &=& \sum_{k=1}^{L} \left(2 |c_k|^2-c_k \bar{c}_{k-1}-\bar{c}_k c_{k-1}
-\frac{\alpha}{2} |c_k|^4\right)\nonumber \\
&=&4- \frac{\alpha}{2} \sum_{k=1}^L |c_k|^4 -\sum_{k=1}^L |\bar{c}_{k-1}+c_k|^2
\nonumber\\
&=&-\frac{\alpha}{2}\sum_{k=1}^L |c_k|^4+ \sum_{k=1}^L |\bar{c}_{k-1}-c_k|^2 \, .
\end{eqnarray}
Furthermore, the so-called inverse participation ratio (which is a measure of localisation) is bounded
\begin{equation}\label{b.3}
\frac{1}{L} \leq \sum_{k=1}^L |c_k|^4 \leq 1
\end{equation}
because
\begin{equation}\label{b.4}
1=\left(\sum_{k=1}^L |c_k|^2\right)^2 =\sum_{k,\ell=1}^L |c_k|^2 |c_\ell|^2
\geq \sum_{k=1}^L |c_k|^2 |c_k|^2
\end{equation}
and by the Cauchy-Schwarz inequality
\begin{equation}\label{b.5}
1=\left(\sum_{k=1}^L |c_k|^2 \cdot 1 \right)
\leq \sum_{k=1}^L |c_k|^4 \cdot L \, .
\end{equation}
To obtain the phase diagram for our model, we determine  the upper and lower energy bounds and the mean energy for $T=\infty$. 
We will discuss the two cases $\alpha>0$ and $\alpha<0$ separately.

{\em Case $\alpha > 0$}: Eq.~\eqref{b.2} and the inequality \eqref{b.3}
results in the upper bound
\begin{equation}\label{b.6}
H_S=4-\frac{\alpha}{2} \sum_{k=1}^L |c_k|^4 -\sum_{k=1}^L |\bar{c}_{k-1}+c_k|^2
\leq 4-\frac{\alpha}{2L} = E_{\mathrm{up}}(\alpha) \, .
\end{equation}    
This upper bound is sharp since for the antiferromagnetic state $c_k=(-1)^k/\sqrt{L}$
eq.~\eqref{b.6} results in $H_S=4-\alpha/(2L)=E_{\mathrm{up}}(\alpha)$. In the  
same vein, eq.~\eqref{b.2} and the inequality \eqref{b.3} result in a lower bound
\begin{equation}\label{b.7}
H_S=-\frac{\alpha}{2}\sum_{k=1}^L |c_k|^4+ \sum_{k=1}^L |\bar{c}_{k-1}-c_k|^2
\geq-\frac{\alpha}{2}
\end{equation}
so that a lower bound $E_{\mathrm{low}}(\alpha)$ exists. The bound stated in
eq.~\eqref{b.7} is unlikely to be sharp, as that would require the state to be
(somehow) ferromagnetic, putting the upper bound for the participation ratio
in eq.~\eqref{b.3} out of reach. In fact, for $c_k=1/\sqrt{L}$ eq.~\eqref{b.7} results in
$H_S=-\alpha/(2L)$, while the localised state $c_k=\delta_{k,k_0}$ gives
$H_S=2-\alpha/2$. Thus, the lower bound of the energy obeys
\begin{equation}\label{b.8}
-\frac{\alpha}{2}\leq E_{\mathrm{low}}(\alpha)\leq \min\left\{-\frac{\alpha}{2L},
2-\frac{\alpha}{2}\right\}
\end{equation}
To clarify the nature of the lower bound, recall that eq.~\eqref{2.1} becomes minimal
if the coordinates $c_k$ are real and non-negative, since the interaction term
$-c_k\bar{c}_{k-1}-c_k\bar{c}_{k+1}$ can be decreased (without affecting the normalisation $N=1$)
by adjusting complex arguments
until all coordinates are real and non-negative. Hence, to determine the lower bound one
needs to solve a minimisation problem on $\mathbb{R}^L$ subjected to a constraint namely
\begin{eqnarray}\label{b.9}
H_S&=&\sum_{k=1}^L\left( 2 r_k^2 - r_k r_{k-1} - r_k r_{k+1} -\frac{\alpha}{2} r_k^4
\right) \rightarrow \mbox{min.}, \sum_{k=1}^L r_k^2=1 \, .
\end{eqnarray}
Because the constraint minimisation takes place over a compact set without boundary, the problem can be strictly solved by Lagrange multipliers (extrema do not occur
on the boundary of the domain), resulting in the condition
\begin{equation}\label{b.10}
(2-\mu) r_k-r_{k-1}-r_{k+1} - \alpha r_k^3=0
\end{equation}
where $\mu$ denotes the Lagrange multiplier. This condition coincides with the
equation for the breather solution of the DNSE (see, for instance 
\cite{EBRAHIMI2025134905}) where the value of the Lagrange multiplier determines the oscillation frequency of the breather. 
Any change of sign of $r_k$ does not alter the constraint or the local interaction term
$\sum_k r_k^4$. Thus, the global minimum is that particular local minimum such that the interaction term $\sum_k r_k r_{k-1}$ becomes maximal, i.e. a profile where neighbouring sites have amplitudes of the same sign.
Thus, the lower bound, i.e., the ground state of the Hamiltonian
is determined by a particular localised breather with a non-negative profile.

\begin{figure}[h!]
    \centering
   \includegraphics[width=0.7\linewidth]{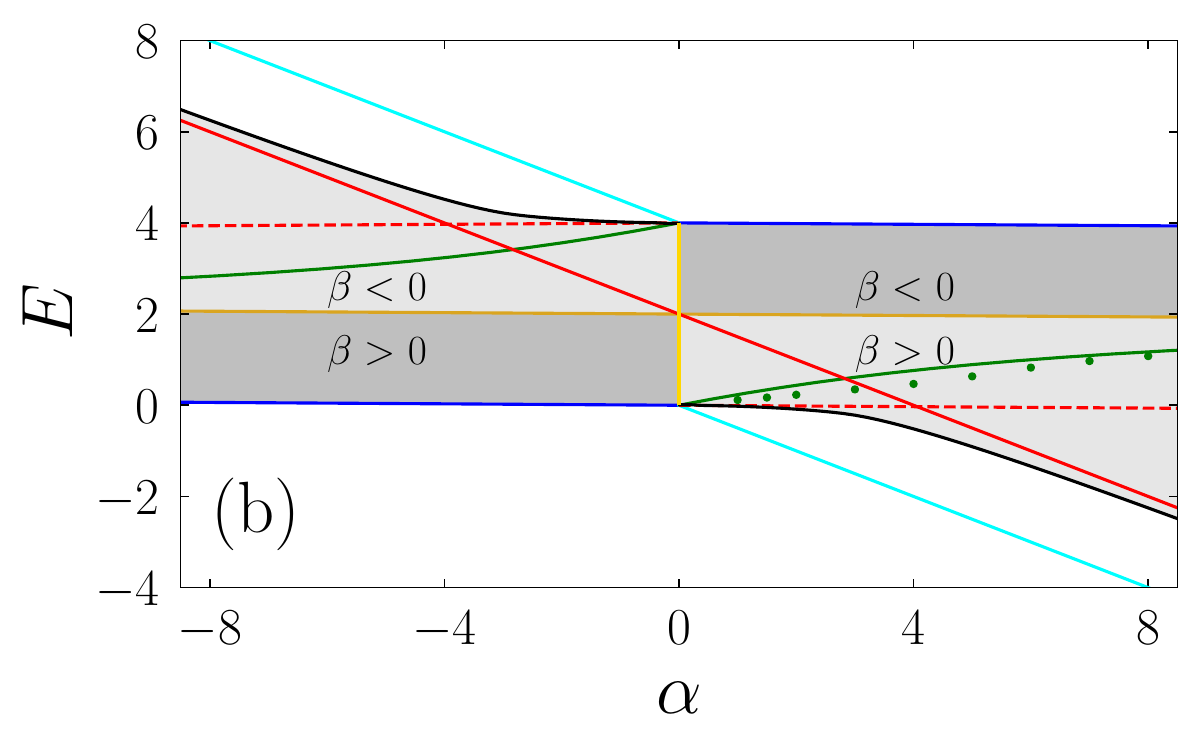}
    \caption{Phase diagram of accessible energies $E$ and on-site
    potential strengths $\alpha$ (shaded regions) for the Hamiltonian 
    \eqref{2.1} with system 
    size $L = 64$. Blue: boundaries $E_{up}(\alpha)$ for $\alpha>0$
    (see eq.~\eqref{b.6}) and $E_{\mathrm{low}}(\alpha)$ for $\alpha<0$
    (see eq.~\eqref{b.12}) caused by ferromagnetic and antiferromagnetic states.
    Black: boundaries $E_{\mathrm{low}}(\alpha)$ for $\alpha>0$ caused by 
    positive breather and $E_{\mathrm{up}}(\alpha)$ for $\alpha<0$
    caused by a staggered breather (this solution was calculated by solving the minimisation problem, eq.~\eqref{b.9}, by numerical means). Cyan and red (solid/dashed): analytic upper
    and lower bounds for the boundary caused by breather states (see 
    eq.~\eqref{b.8} for $\alpha>0$ and eq.~\eqref{b.14} for $\alpha<0$).
    Bronze: infinite temperature line (see eq.~\eqref{b.11}). 
    Light shading                        
    indicates the region of predominantly localised breather-type states, dark shading      
    the region of predominantly antiferromagnetic ($\alpha>0$) or    
    ferromagnetic ($\alpha<0$) states.    
    The different types
    of shading also correlates with positive and negative temperature regions
    (as indicated by labels). Yellow: the line $\alpha=0$ for which an
    explicit analytic expression for the caloric equation of state, $E(\beta)$, is
    available (see eq.~\eqref{e.16}). Green: Phase transition line, eq.~\eqref{4.5a}, obtained by 
    mean field theory (see section \ref{sec:4} for details).
    The green dots are numerical results for the breakdown of the localised phase
    obtained from simulations in a system of size $L=1024$ with noise strength $\sigma=0.3$,
    using time series of length $6.8\times 10^5$.}
    \label{figb1}
\end{figure}

The ground state of the Hamiltonian with energy 
$E_{\mathrm{low}}(\alpha)$ corresponds to the zero-temperature state,
$\beta=\infty$. Since our Hamiltonian is bounded, the state with
highest energy $E_{\mathrm{up}}(\alpha)$ corresponds to 
the high-temperature limit $\beta=-\infty$ which means a negative-temperature state
where one approaches zero temperature from below. Such features are quite
common in systems with bounded energy, e.g., spin systems or Laser systems
(population inversion is a negative-temperature state, which is unstable 
in experiments since atoms are coupled to a positive temperature, i.e., colder heat bath), 
but they do not occur in conventional mechanical systems, as the kinetic 
energy is unbounded. In the latter, the highest-temperature state corresponds 
to $\beta=0$, i.e., the positive temperature tending to infinity.
In the setup of the DNSE with conserved normalisation $N=1$ 
the infinite-temperature state $\beta=0$ is intermediate 
between the low-temperature state $\beta=\infty$ and the high-temperature state
$\beta=-\infty$. The corresponding distribution in phase space
is uniform on the sphere determined by the conservation law $N=1$,
see eq.~\eqref{2.1.3}, and the corresponding expectation values can be
easily computed (see for instance \cite{EBRAHIMI2025134905}) to be
$\langle |c_k|^2\rangle=1/L$, $\langle |c_k|^4\rangle=2/(L(L+1))$, 
and $\langle c_k \bar{c}_{k-1}\rangle=0$. The expectation of the 
Hamiltonian \eqref{b.2} then defines 
the infinite-temperature line in the phase diagram
\begin{equation}\label{b.11}
E_\infty(\alpha)=\langle H_S\rangle=2-\frac{\alpha}{L+1} \, .
\end{equation}
Energies below this value correspond to positive --, energies above this value
to negative temperatures (see Figure \ref{figb1}).

{\em Case $\alpha<0$}: We can deal with this case either by employing the
symmetry discussed in section \ref{sec:2.2} or by a simple calculation
along the lines of the previous case.

Eqs.~\eqref{b.2} and \eqref{b.3} give the lower bound
\begin{equation}\label{b.12}
H_S =\frac{(-\alpha)}{2}\sum_{k=1}^L |c_k|^4+ \sum_{k=1}^L |\bar{c}_{k-1}-c_k|^2 
\geq -\frac{\alpha}{2 L} = E_{\mathrm{low}}(\alpha)
\end{equation}
and this lower bound is sharp since for a ferromagnetic state $c_k=1/\sqrt{L}$
eq.~\eqref{b.12} results in $H_S=-\alpha/(2L)$. Similarly, eqs.~\eqref{b.2}
and \eqref{b.3} yield the upper bound
\begin{equation}\label{b.13}
H_S=4 +\frac{(-\alpha)}{2} \sum_{k=1}^L |c_k|^4 -\sum_{k=1}^L |\bar{c}_{k-1}+c_k|^2
\leq 4-\frac{\alpha}{2}
\end{equation}
and this upper bound is not sharp. For a localised state $c_k=\delta_{k,k_0}$
eq.~\eqref{2.1} gives $H_S=2-\alpha/2$ while a for an antiferromagnetic state
$c_k=(-1)^k/\sqrt{L}$ eq.~\eqref{b.13} results in $H_S=4-\alpha/(2L)$. Thus, the
sharp upper bound $E_{\mathrm{up}}(\alpha)$ obeys
\begin{equation}\label{b.14}
\max\left\{
4-\frac{\alpha}{2L},2-\frac{\alpha}{2}
\right\}\leq E_{\mathrm{up}}(\alpha) \leq 4-\frac{\alpha}{2} \, .
\end{equation}
As before, the exact value of the upper bound follows by maximising \eqref{2.1}
subjected to the constraint $N=1$. Since the interaction term in the Hamiltonian 
becomes maximal for real coordinates with alternating signs, we thus obtain a
maximisation problem analogous to eq.~\eqref{b.9}. The condition for
the maximal state is again given by eq.~\eqref{b.10} with a real solution
where the sign alternates between adjacent lattice sites. Such a solution is
generally accepted to be the staggered breather solution of the DNSE discussed, for instance, in
\cite{EBRAHIMI2025134905}. Finally, for the infinite-temperature line eq.~\eqref{b.11} 
applies as this part of the previous consideration 
did not involve the sign of $\alpha$.

In summary, these considerations can be condensed into a phase diagram
which shows the accessible region in the $E$-$\alpha$ parameter plane together
with positive and negative temperature domains (see figure \ref{figb1}).
The phase diagram displays the symmetry mentioned in section \ref{sec:2.2}.

\section{Fokker-Planck equation and stationary distribution}\label{sec:c}

The dynamics of the stochastic differential equation \eqref{2.2} can be captured by
the probability distribution of phase space coordinates $\rho_t(\{c_k,\bar{c}_k\})$
which obeys the Fokker-Planck equation \cite{Risk:89}

\begin{equation}\label{c.1}
\frac{\partial \rho_t}{\partial t}
=\sum_{k=1}^L \left(
-\frac{\partial}{\partial c_k} D^{(1)}_{c_k}
-\frac{\partial}{\partial \bar{c}_k} D^{(1)}_{\bar{c}_k}
+\frac{\partial^2}{\partial c_k^2} D^{(2)}_{c_k c_k}
+2 \frac{\partial^2}{\partial c_k \partial \bar{c}_k} D^{(2)}_{c_k \bar{c}_k}
+\frac{\partial^2}{\partial \bar{c}_k^2} D^{(2)}_{\bar{c}_k \bar{c}_k}
\right) \rho_t
\end{equation}

with drift and diffusion coefficients being given by, following the Stratonovich rules
\begin{eqnarray}\label{c.2}
D^{(1)}_{c_k} &=& i \frac{\partial H_S}{\partial \bar{c}_k} -
\beta \frac{\sigma^2}{2} c_k \left( \bar{c}_k
\frac{\partial H_S}{\partial \bar{c}_k}
- c_k \frac{\partial H_S}{\partial c_k}
\right) -\frac{\sigma^2}{2} c_k, \nonumber \\
D^{(1)}_{\bar{c}_k} &=& \overline{D^{(1)}_{c_k}}, \qquad
D^{(2)}_{c_k c_k} =-\frac{\sigma^2}{2} c_k^2, \nonumber \\
D^{(2)}_{c_k \bar{c}_k} &=& \frac{\sigma^2}{2} |c_k|^2, \qquad
D^{(2)}_{\bar{c}_k \bar{c}_k} = \overline{D^{(2)}_{c_k c_k}}
\end{eqnarray}
With a little bit of algebra eqs.(\ref{c.1}) and (\ref{c.2}) can be written as
\begin{equation}\label{c.3}
\frac{\partial \rho_t}{\partial t} = -\{H_S,\rho_t\}
-\sum_{k=1}^L \left(
\frac{\partial J_{c_k}(\rho_t)}{\partial c_k}
+ \frac{\partial J_{\bar{c}_k}(\rho_t)}{\partial \bar{c}_k}
\right)
\end{equation}
where
\begin{eqnarray}\label{c.4}
J_{c_k}(\rho_t) &=& \frac{\sigma^2}{2} c_k^2
\left(\beta \frac{\partial H_S}{\partial c_k} \rho_t+\frac{\partial \rho_t}{\partial c_k}
\right)-
\frac{\sigma^2}{2} |c_k|^2
\left(\beta \frac{\partial H_S}{\partial \bar{c}_k} \rho_t+
\frac{\partial \rho_t}{\partial \bar{c}_k}
\right)\nonumber \\
J_{\bar{c}_k}(\rho_t) &=& \overline{J_{c_k}(\rho_t)} 
\end{eqnarray}
denote the so-called irreversible currents. Obviously $\exp(-\beta H_S)$ is 
a (potentially not normalisable) stationary solution
since $\{ H_S,\exp(-\beta H_S)\}=0$ and the irreversible currents vanish,
$J_{c_k}(\exp(-\beta H_S))=0$. That means that the stationary solution obeys detailed
balance.

If $\rho_t$ is a solution of the Fokker-Planck equation \eqref{c.3} then
$\rho_t f(N)$ is a solution as well for any (reasonable) function $f$ which only depends
on the normalisation \eqref{2.1a}. To see this note that on the one hand
$\{H_S,\rho_t f(N)\}=f(N) \{H_S, \rho_t\}$. On the other hand, we conclude from
\begin{eqnarray}\label{c.5}
c_k^2\frac{\partial \rho_t f(N)}{\partial c_k}
-|c_k|^2\frac{\partial \rho_t f(N)}{\partial \bar{c}_k} &=& f(N)\left(c_k^2\frac{\partial \rho_t}{\partial c_k}
-|c_k|^2\frac{\partial \rho_t}{\partial \bar{c}_k}\right)
 + \rho_t f'(N)\left( c_k^2 \bar{c}_k-|c_k|^2 c_k\right) \nonumber \\
&=&
f(N)\left(c_k^2\frac{\partial \rho_t}{\partial c_k}
-|c_k|^2\frac{\partial \rho_t}{\partial \bar{c}_k}\right)
\end{eqnarray}
that the definition \eqref{c.4} implies
\begin{equation}\label{c.6}
J_{c_k}(\rho_t f(N))=f(N) J_{c_k}(\rho_t) \, .
\end{equation}
Using 

\begin{eqnarray}\label{c.7}
\bar{c}_k J_{c_k}(\rho_t)+c_k J_{\bar{c}_k}(\rho_t)
&=&\frac{\sigma^2}{2} \left(\beta \frac{\partial H_S}{\partial c_k}
+\frac{\partial \rho_t}{\partial c_k}\right)
\left(\bar{c}_k c_k^2 - c_k |c_k|^2\right) \nonumber\\&&+\frac{\sigma^2}{2}
\left(\beta \frac{\partial H_S}{\partial \bar{c}_k}
+\frac{\partial \rho_t}{\partial \bar{c}_k}\right)
\left(c_k \bar{c}_k^2-\bar{c}_k |c_k|^2 \right) =0
\end{eqnarray}

we therefore have 

\begin{eqnarray}\label{c.8}
\frac{\partial J_{c_k}(\rho_t f(N))}{\partial c_k}+
\frac{\partial J_{\bar{c}_k}(\rho_t f(N))}{\partial \bar{c}_k}
&=& f(N)\left( 
\frac{\partial J_{c_k}(\rho_t)}{\partial c_k}+
\frac{\partial J_{\bar{c}_k}(\rho_t)}{\partial \bar{c}_k}
\right)
+\rho_t f'(N) \left(\bar{c}_k J_{c_k}(\rho_t)+ c_k J_{\bar{c}_k}(\rho_t)\right)
\nonumber \\
&=&f(N)\left( 
\frac{\partial J_{c_k}(\rho_t)}{\partial c_k}+
\frac{\partial J_{\bar{c}_k}(\rho_t)}{\partial \bar{c}_k}
\right) \, .
\end{eqnarray}

Altogether that means eq.~\eqref{c.3} implies that
\begin{eqnarray}\label{c.9}
\frac{\partial \rho_t f(N)}{\partial t} &=& -f(N) \{H_S,\rho_t\} -f(N) \sum_{k=1}^L \left(
\frac{\partial J_{c_k}(\rho_t)}{\partial c_k}
+ \frac{\partial J_{\bar{c}_k}(\rho_t)}{\partial \bar{c}_k}
\right)\nonumber \\
&=& -\{H_S,\rho_t f(N)\} -\sum_{k=1}^L \left(
\frac{\partial J_{c_k}(\rho_t f(N))}{\partial c_k}
+ \frac{\partial J_{\bar{c}_k}(\rho_t f(N))}{\partial \bar{c}_k}
\right)
\end{eqnarray}
so $\rho_t f(N)$ is a solution of the Fokker-Planck equation for any function $f$. In particular
\begin{equation}\label{c.10}
\rho_\beta=\exp(-\beta H_S) f(N)
\end{equation}
is a stationary solution which is normalisable if, for instance, 
$f$ has compact support. Eq.~\eqref{2.1.3} is an installment of eq.~\eqref{c.10} 
when the dynamics are constrained by the condition $N=1$.

Finally, eq.~\eqref{c.3} and the stationary solution \eqref{c.10} admit an H-theorem.
For that purpose we adjust the standard reasoning, 
see for instance \cite{Risk:89}, and introduce a relative entropy $L(t)$,
sometimes called the Kullback-Leibler divergence, which will serve 
as a Lyapunov function.
Consider the expression
\begin{equation}\label{c.11}
L(t)=\int_{\mathbb{C}^L} e^{-\beta H_S} \ln\left(\frac{e^{-\beta H_S}}{\rho_t }\right)
f(N) \prod_{k=1}^L d(c_k,\bar{c}_k) \, .
\end{equation}
where we impose the normalisation $\int_{\mathbb{C}^L} \rho_\beta d(c_k,\bar{c}_k)
= \int_{\mathbb{C}^L} \rho_t f(N) d(c_k,\bar{c}_k)$. With the standard inequality
$x\ln x-x+1>0$ if $x>0$ and $x\neq 1$ we conclude that
\begin{align}\label{c.12}
L(t)
=\int_{\mathbb{C}^L}
\Big[
    \frac{e^{-\beta H_S}}{\rho_t}
    \ln\!\frac{e^{-\beta H_S}}{\rho_t}
    - \frac{e^{-\beta H_S}}{\rho_t}
    + 1
\Big]
\rho_t\, f(N)
\prod_{k=1}^L d(c_k,\bar{c}_k)
\geq 0 .
\end{align}
In fact $L(t)>0$ if $\rho_t f(N)\neq \rho_\beta$ and
$L(t)=0$ if $\rho_t f(N) = \rho_\beta$ so that $L(t)$ 
has a global minimum at $\rho_t f(N) =\rho_\beta$. Furthermore 

\begin{eqnarray}\label{c.13}
\dot{L}(t)
&=&- \int_{\mathbb{C}^L}
\frac{e^{-\beta H_S}}{\rho_t} \left(
-\{H_S, \rho_t\} -\sum_{k=1}^L \left(\frac{\partial J_{c_k}(\rho_t)}{\partial c_k}
+\frac{\partial J_{\bar{c}_k}(\rho_t)}{\partial \bar{c}_k}
\right)
\right) \times f(N) \prod_{k=1}^L d(c_k,\bar{c}_k)\nonumber \\
&=& \int_{\mathbb{C}^L} e^{-\beta H_S} \{H_S, \ln \rho_t\} f(N) 
\prod_{k=1}^L d(c_k,\bar{c}_k) \nonumber \\&&+ \sum_{k=1}^L \int_{\mathbb{C}^L} \frac{e^{-\beta H_S}}{\rho_t} 
\left(\frac{\partial J_{c_k}(\rho_t)}{\partial c_k}
+\frac{\partial J_{\bar{c}_k}(\rho_t)}{\partial \bar{c}_k}
\right) f(N) 
\prod_{k=1}^L d(c_k,\bar{c}_k)\nonumber \\
&=& - \sum_{k=1}^L \int_{\mathbb{C}^L} 
\left(\frac{\partial e^{-\beta H_S}/\rho_t}{\partial c_k} J_{c_k}(\rho_t) + \frac{\partial e^{-\beta H_S}/\rho_t}{\partial \bar{c}_k} J_{\bar{c}_k}(\rho_t)
\right) \times f(N) 
\prod_{k=1}^L d(c_k,\bar{c}_k)
\end{eqnarray}

where we have used eq.~\eqref{c.7} to perform the
final integration by parts. Since the irreversible currents, eq.~(\ref{c.4}),
can be written as
\begin{equation}\label{c.14}
J_{c_k}(\rho_t)=\frac{\sigma^2}{2}e^{-\beta H_S}\left(
c_k^2 \frac{\partial \rho_t/e^{-\beta H_S}}{\partial c_k}
-|c_k|^2 \frac{\partial \rho_t/e^{-\beta H_S}}{\partial\bar{c}_k}
\right)
\end{equation}
eq.~\eqref{c.13} results in

\begin{equation}\label{c.15}
\dot{L}(t)=-\frac{\sigma^2}{2}\sum_{k=1}^L
\int_{\mathbb{C}^L} \left(\frac{e^{-\beta H_S}}{\rho_t}\right)^2
(\bar{p}_k,p_k) \left(
\begin{array}{rr} 
1 & -1 \\
-1 & 1
\end{array}
\right) \left(
\begin{array}{c}
p_k \\ \bar{p}_k
\end{array}
\right) e^{-\beta H_S}f(N) 
\prod_{k=1}^L d(c_k,\bar{c}_k)
\end{equation}

where we have employed the abbreviation 
\begin{equation}\label{c.16}
p_k =c_k \frac{\partial \rho_t/e^{-\beta H_S}}{\partial c_k} \, .
\end{equation}
Since the quadratic form in the integrals of eq.~\eqref{c.15} is positive semi-definite
we have $\dot{L}(t)\leq 0$, as expected from a Lyapunov function. Thus, the dynamics
of the Fokker-Planck equation results in a monotonic decrease of $L(t)$ until we reach
$\dot{L}(t)=0$. That of course occurs if $\rho_t f(N)=\rho_\beta$. Conversely, the condition 
$\dot{L}(t)=0$ requires all the quadratic forms in eq.~\eqref{c.15} to
vanish for any values of $\{c_k\}$, that means $p_k=\bar{p}_k$ is real. That requires
the ratio $\rho_t/e^{-\beta H_S}$ to be a function of the absolute values only, i.e.
$\rho_t/e^{-\beta H_S}=g(\{|c_k|^2\},t)$. If we plug this expression into the equation of motion \eqref{c.3} a short calculation along the lines mentioned above shows that the
irreversible currents do not contribute and we are left with $\partial g/\partial t=
-\{H_S,g\}$. Since the Hamiltonian \eqref{2.1} is unlikely to have any further
conserved quantity depending only on $\{ |c_k|^2\}$, apart from the normalization \eqref{2.1a} we
are left with the time independent choice that depends on $N$ only,
$g(\{ |c_k|^2,t\}=h(N)$. Thus, the dynamics of the Fokker-Planck equation equilibrate
on any shell $N=c\geq 0$ but not between different shells. If the dynamics is
constrained by the condition $N=1$ we hence obtain a unique stationary state,
eq.~\eqref{2.1.3}, for systems of finite size. As usual, the relaxation times to
reach this equilibrium may be huge, depending on the value of the system size.
It is finally worth mentioning that the validity of the H-theorem does not require the temperature to be positive.

\section{Numerical integration scheme}\label{sec:d}

The SDNSE, eq. \eqref{2.2.1}, preserves normalisation $N$. To avoid numerical
artefacts a numerical integration scheme should preserve the
normalisation as well. Furthermore, energy is conserved if
the dissipation and the noise vanish.
The numerical integration scheme should become a symplectic scheme
in the limit $\sigma\rightarrow 0$ to have uniformity in $\sigma$.
These conditions put
some constraints on suitable numerical integration schemes. It
is quite well known how to cope with these issues, and here we
adjust sympletic integration schemes developed e.g. in \cite{YOSHIDA1990262}.
We by and large follow the exposition described previously in \cite{EBRAHIMI2025134905}
and we focus on the required adjustment in the dissipative context.
The evolution operator of eq.~\eqref{2.2.1} acting on a function $F$
is defined by $F(\{c_k(t),\bar{c}_k(t)\})=\mathbf{U}(t_0,t) F(\{c_k(t_0),\bar{c}_k(t_0)\})$. 
The operator obeys the usual linear differential equation
\begin{eqnarray}
\label{d.1.2}
&&\frac{\partial \mathbf{U}(t_0,t)}{\partial t}
=\mathbf{U}(t_0,t) \left(i\mathbf{L}_A + i\mathbf{L}_B(t)
- \mathbf{\Lambda}\right) , \qquad \mathbf{U}(t_0,t_0)=\mathbf{1}
\end{eqnarray}
where $i\mathbf{L}_A.=\{H_A,.\}$ and $i \mathbf{L}_B(t).=\{H_B(t),.\}$
denote Liouville operators associated with Hamiltonian parts of the motion
\begin{eqnarray} \label{d.1.1}
H_A &=& \sum_{n=1}^L\left(2|c_n|^2 - c_n\bar{c}_{n-1} - c_n\bar{c}_{n+1}\right)
\nonumber \\
H_B(t) &=& \sum_{n=1}^L
\left(-\frac{\alpha}{2}|c_n|^4 + \sigma |c_n|^2\xi_n(t) \right)
\end{eqnarray}
and $\mathbf{\Lambda}$ captures the dissipative contributions

\begin{eqnarray}\label{d.1.1a}
\mathbf{\Lambda} F &=& \beta \frac{\sigma^2}{2} \sum_{k=1}^L c_k \left(
c_k(\bar{c}_{k+1}+\bar{c}_{k-1})-\bar{c}_k(c_{k+1}+c_{k-1})
\right)\frac{\partial F}{\partial c_k}
\nonumber \\ &&+ \beta \frac{\sigma^2}{2} \sum_{k=1}^L \bar{c}_k \left(
\bar{c}_k (c_{k+1}+c_{k-1}) - c_k(\bar{c}_{k+1}+\bar{c}_{k-1})
\right) \frac{\partial F}{\partial \bar{c}_k} \, .\nonumber\\
\end{eqnarray}

The evolution operator can be expressed as a time ordered
exponential. As shown, for instance, in \cite{EBRAHIMI2025134905} this exponential
can be formally written as a product, down to second order terms in the stepsize $\tau$
\begin{eqnarray}
\label{d.1.5}
\mathbf{U}(t_0,t_0+\tau) =
\exp\left(i \mathbf{L}_B^<\tau \right)
\exp\left(i \mathbf{L}_A \tau-\mathbf{\Lambda}\tau\right)
\exp\left(i \mathbf{L}_B^>\tau\right)+ \mathcal{O}(\tau^3)
\end{eqnarray}
where $i \mathbf{L}_B^\lessgtr . =\{H_B^\lessgtr,.\}$ represent Liouville
operators of effective time independent Hamiltonians
\begin{equation}\label{d.1.6}
H_B^\lessgtr= \frac{1}{2} \sum_{n=1}^{L}
\left( -\frac{\alpha}{2}\right) |c_n|^4
+ \sum_{n=1 }^{L} \sigma |c_n|^2 \frac{D_n^\lessgtr}{\sqrt{\tau}}  \, .
\end{equation}
Here, $D_n^\lessgtr$ denote Gaussian random numbers with correlations
\begin{eqnarray}\label{d.1.8}
&&\langle D_n^< D_m^<\rangle = \frac{1}{3} \delta_{n,m}, \qquad
\langle D_n^> D_m^>\rangle = \frac{1}{3} \delta_{n,m}, \qquad
\langle D_n^< D_m^>\rangle = \frac{1}{6} \delta_{n,m} \, .
\end{eqnarray}
Using, for instance, the usual series expansion, we can finally write the
combined exponential in eq.~\eqref{d.1.5} as a product,
down to terms of second order
\begin{eqnarray}\label{d.1.8a}
\exp\left(i \mathbf{L}_A \tau -\mathbf{\Lambda}\tau\right)
= \exp\left(-\mathbf{\Lambda}\tau/2\right)
\exp\left(i \mathbf{L}_A \tau\right)
\exp\left(-\mathbf{\Lambda}\tau/2\right)+ \mathcal{O}(\tau^3) \, .
\end{eqnarray}
Eqs.~\eqref{d.1.5} and \eqref{d.1.8a} define a numerical 
integration step with stepsize $\tau$,
where each integration step consists of five individual transformations.
The action of the exponentials $\exp(i \mathbf{L}_A \tau)$ and
$\exp(i \mathbf{L}_B^\lessgtr \tau)$ can be calculated in closed analytic form,
as the corresponding dynamics is governed by a system of independent
entities or by a nearest neighbour coupled harmonic chain, cf.
eqs.~\eqref{d.1.1} and \eqref{d.1.6} (see \cite{EBRAHIMI2025134905} for 
the explicit analytic expressions).
In addition, those transformations are symplectic and they preserve
the normalisation so that these parts meet the constraints mentioned above.
We still need to discuss the effect of the dissipative part
$\exp(-\mathbf{\Lambda} \tau/2)$.

For that purpose define $c_n(t)=\exp(-\mathbf{\Lambda} t) c_n$.
Using eq.\eqref{d.1.1a} these functions obey the system
of differential equations
\begin{align}\label{d.1.8b}
\dot{c}_n(t)
= -\beta \frac{\sigma^2}{2} c_n(t)
\Big[
  c_n(t)\big( \bar{c}_{\,n-1}(t) + \bar{c}_{\,n+1}(t) \big)
  - \bar{c}_n(t)\big( c_{\,n-1}(t) + c_{\,n+1}(t) \big)
\Big].
\end{align}
By construction, this set of equations just contains the dissipative part
of the original SDNSE, eq.~\eqref{2.2.1}. That means we are now going to discuss the
sole impact of the damping on the motion. It is fairly straightforward to show
that eq.~\eqref{d.1.8b} preserves the weights $|c_n|^2$ since

\begin{eqnarray}
\label{d.2.1}
\frac{d |c_n|^2}{d t} &= &
\dot{c}_n(t)\bar{c}_n(t) + c_n(t) \dot{\bar{c}}_n(t) \nonumber\\
&= &-\beta\frac{\sigma^2}{2}
|c_n(t)|^2\Big[ c_n(t)(\bar{c}_{n-1}(t) + \bar{c}_{n+1}(t))
- \bar{c}_n(t) (c_{n-1}(t) + c_{n+1}(t))\Big] \nonumber\\
& & - \beta\frac{\sigma^2}{2} |c_n(t)|^2\Big[
\bar{c}_n(t)(c_{n-1}(t) + c_{n+1}(t)) - c_n(t)
(\bar{c}_{n-1}(t)+ \bar{c}_{n+1}(t)) \Big] = 0 \, .
\end{eqnarray}

Such a property already implies that the action of
$\exp(-\mathbf{\Lambda} \tau/2)$ preserves the
normalisation since each individual weight
$|c_n ^2|$ remains constant.
In addition, we can conclude
that the time dependent solutions of eq.~\eqref{d.1.8b}
have a constant absolute value
\begin{equation}
\label{d.2.3}
c_n(t) = |c_n| e^{i\theta_n(t)}
\end{equation}
where the time dependent complex phases obey (see eq.~\eqref{d.1.8b})
\begin{align}\label{d.2.5}
\dot{\theta}_n(t)
= \beta\sigma^2 |c_n|\Big[
  |c_{n-1}|\sin\!\left(\theta_{n-1}(t)-\theta_n(t)\right)
  + |c_{n+1}|\sin\!\left(\theta_{n+1}(t)-\theta_n(t)\right)
\Big].
\end{align}
Hence, the damping mechanism on its own only readjusts the complex
phases but leaves the individual $|c_n|$ untouched.
Eq.~\eqref{d.2.5} has the form of a phase coupled oscillator model, like
the Kuramoto model, with nearest neighbour couplings determined by the
constant values of the $|c_n|$. While no analytic closed-form solution is available
in these cases, we may compute approximate values for $\theta_n(\tau/2)$
by a single numerical integration step of second order accuracy, e.g., by a step
of the midpoint method. Even with these numerical errors the resulting values
for $c_n(\tau/2)=\exp(-\mathbf{\Lambda}\tau/2)c_n =
|c_n| \exp(i\theta_n(\tau/2))$ will preserve normalisation, so that our scheme
fulfils all the requirements mentioned above.

\section{The harmonic chain}\label{sec:e}

For $\alpha=0$ the system \eqref{2.1} reduces to the discrete linear
Schr\"odinger equation, or the harmonic chain, where closed analytic
expressions become available. For the computation of the corresponding partition
integral \eqref{2.1.4} we just have to observe the global constraint caused by
the $\delta$-contribution. Hence, the calculation and the result differ
slightly from the usual canonical expressions.

Using Fourier modes
\begin{equation}\label{e.1}
\hat{c}_q = \frac{1}{\sqrt{L}}\sum_{k=1}^L e^{i q k} c_k
\end{equation}
with wavenumbers $q=q_\nu=2\pi \nu/L$, $\nu=0,1,\ldots,L-1$
we have as usual, using eq.~\eqref{2.1} with $\alpha=0$ and eq.~\eqref{2.1a}

\begin{eqnarray}\label{e.2}
N &=& \sum_{\nu=0}^{L-1} \left| \hat{c}_{q_\nu}\right|^2\nonumber\\
H_S&=&\sum_{\nu=0}^{L-1} 2(1-\cos(q_\nu)) \left| \hat{c}_{q_\nu}\right|^2
=\sum_{\nu=1}^{L-1} \omega_{q_\nu}^2 \left| \hat{c}_{q_\nu}\right|^2
\end{eqnarray}
where $\omega_q=\sqrt{2(1-\cos(q)}$ denotes the dispersion relation of the
linear chain, and the mode $q=0$ drops from the Hamiltonian. Therefore we can
integrate out this mode and the partition integral \eqref{2.1.4} reads
\begin{eqnarray}\label{e.3}
Z_\beta = \pi^L \int_{\mathbb{R}_+^{L-1}} 
\exp\left(-\beta \sum_{\nu=1}^{L-1} \omega^2_{q_\nu} \left| 
\hat{c}_{q_\nu}\right|^2 \right)
\Theta\left(
1-\sum_{\nu=1}^{L-1}  \left| 
\hat{c}_{q_\nu}\right|^2 \right) \prod_{\nu=1}^{L-1} d
\left| 
\hat{c}_{q_\nu}\right|^2
\end{eqnarray}
where $\Theta(x)$ denotes the Heaviside step function. To isolate the
temperature dependence, we introduce new, positive integration variables
$y_\nu=|\beta| \left| \hat{c}_{q_\nu}\right|^2$ as well as the sign of the
temperature by $\eta=\mbox{sgn}(\beta)$. Then eq.~\eqref{e.3} reads
\begin{equation}\label{e.4}
Z_\beta =\frac{\pi^L}{|\beta|^{L-1}} \zeta(|\beta|)
\end{equation}
where we have used the abbreviation
\begin{eqnarray}\label{e.5}
\zeta(|\beta|) = \int_{\mathbb{R}_+^{L-1}}
\exp\left(-\eta \sum_{\nu=1}^{L-1} \omega_{q_\nu}^2 y_\nu\right)
\Theta\left(|\beta|-\sum_{\nu=1}^{L-1} y_\nu \right) \prod_{\nu=1}^{L-1} dy_\nu \, .
\end{eqnarray}
Because of the Heaviside function, the range of integration in eq.~\eqref{e.5}
becomes a simplex. Using the somewhat canonical coordinates
\begin{eqnarray}\label{e.6}
y_\nu=t_\nu-t_{\nu+1}, \qquad \nu=1,2,\ldots, L-2, \qquad
y_{L-1}=t_{L-1}
\end{eqnarray}    
the integral can be written as a sequence of convolutions

\begin{eqnarray}\label{e.7}
& &\zeta(|\beta|)=\int_0^{|\beta|} dt_1 \int_0^{t_1} dt_2
e^{-\eta \omega_{q_1}^2 (t_1-t_2)}
\int_0^{t_2} dt_3 e^{-\eta \omega_{q_2}^2 (t_2-t_3)}
\int_0^{t_2} dt_3 \ldots \nonumber \\
& & \int_0^{t_{L-3}} dt_{L-2} e^{-\eta \omega_{q_{L-3}}^2 (t_{L-3}-t_{L-2})}
\int_0^{t_{L-2}} dt_{L-1} e^{-\eta \omega_{q_{L-2}}^2 (t_{L-2}-t_{L-1})}
e^{-\eta \omega_{q_{L-1}}^2 t_{L-1}} \, .
\end{eqnarray} 

Such an expression can be simplified if we employ the Laplace transform
\begin{eqnarray}\label{e.8}
\mathcal{L}(\zeta)(s) &=& \int_0^\infty e^{-s \beta} \zeta(\beta) d\beta
= \frac{1}{s} \prod_{\nu=1}^{L-1} \frac{1}{s+\eta \omega_{q_\nu}^2} \nonumber\\
&=& \exp\left(-\sum_{\nu=0}^{L-1} \ln(s+\eta \omega_{q_\nu^2})\right) \, .
\end{eqnarray}    
In the asymptotic limit of a large system size, the sum in the exponent can be written
as an integral, up to exponentially small terms, i.e.,
\begin{eqnarray}\label{e.9}
\frac{1}{L}\sum_{\nu=0}^{L-1} \ln(s+\eta 2(1-\cos(q_\nu))) 
&\simeq & \frac{1}{2\pi} \int_0^{2 \pi} \ln(s+\eta 2 (1-\cos(q))) dq  \\
&=&
\left\{ \begin{array}{lcr}
\ln\left(\frac{s+2}{2}+ \sqrt{\left(\frac{s+2}{2}\right)^2 -1}\right) \mbox{ if }
\eta=+1, \, \mbox{Re}(s)\geq 0\\\nonumber
\ln\left(\frac{s-2}{2}+ \sqrt{\left(\frac{s-2}{2}\right)^2 -1}\right) \mbox{ if }
\eta=-1, \, \mbox{Re}(s)\geq 4          
\end{array}
\right. \, 
\end{eqnarray}    
If we use this asymptotic result in eq.~\eqref{e.8} and recall that the Laplace transform of the modified Bessel function of the first kind, $I_n(t)$, obeys
\begin{eqnarray}\label{e.10}
\mathcal{L}(n I_n(t)/t)(w)&=&\left(w-\sqrt{w^2-1}\right)^n 
= \frac{1}{\left(w+\sqrt{w^2-1}\right)^n},
\quad \mbox{Re}(w)> 1 
\end{eqnarray}
we finally obtain
\begin{equation}\label{e.11}
\zeta(|\beta|)=e^{-2 \eta |\beta|} \frac{L I_L(2|\beta|)}{|\beta|}
\end{equation}
so that the partition integral, eq.~\eqref{e.4}, reads
\begin{equation}\label{e.12}
Z_\beta=\pi^L e^{-2 \beta} \frac{L I_L(2|\beta|)}{|\beta|^L} \, .
\end{equation}

The standard identities $I_{n-1}(z)-I_{n+1}(z)=2 n I'_n(z)/z$ and
$I_{n-1}(z)+I_{n+1}(z)=2 I'_n(z)$ of the modified Bessel function imply
\begin{equation}\label{e.13}
\frac{d}{dz} \ln \left(\frac{I_n(z)}{z^n}\right)=\frac{I_{n+1}(z)}{I_n(z)}
\end{equation}
and
\begin{equation}\label{e.14}
\frac{I_{n-1}(nx)}{I_n(nx)}-\frac{I_{n+1}(nx)}{I_n(nx)}= \frac{2}{x}
\end{equation}
Thus eqs.~\eqref{e.12} and \eqref{e.13} yield for the caloric equation of state
\begin{eqnarray}\label{e.15}
E(\beta) =-\frac{\partial \ln Z_\beta}{\partial \beta}
= 2 -2 \eta \frac{I_{L+1}(2|\beta|)}{I_L(2|\beta|)}
=2 -2 \eta \frac{I_{L+1}(L\cdot 2|\beta|/L)}{I_L(L \cdot 2|\beta|/L)}
\end{eqnarray}
If we appeal to eq.~\eqref{e.14} in the limit of large $n$, we may conclude that
$I_{n+1}(nx)/I_n(nx)\simeq \alpha(x)$ where $1/\alpha(x)-\alpha(x)=2/x$ so that
$\alpha(x)=x/(1+\sqrt{1+x^2})$. Using that property in eq.~\eqref{e.15} we
obtain the simple analytic expression
\begin{eqnarray}\label{e.16}
E(\beta)&\simeq& 2 -2 \eta \frac{2 | \beta|/L}{1+\sqrt{1+4 \beta^2/L^2}}\nonumber\\
&=&2-2 \frac{2 \beta}{L+\sqrt{L^2+4 \beta^2}} \, .
\end{eqnarray}
The expression fits quite well with numerical results, even for moderate system size
(cf. as well Figure \ref{fig31}(a)). The constraint $N=1$ has resulted in a fairly unusual dependence
of the energy on temperature and system size in this seemingly trivial setup.

\printbibliography

@article{chatterjee2012probabilistic,
  title={Probabilistic methods for discrete nonlinear Schr{\"o}dinger equations},
  author={Chatterjee, Sourav and Kirkpatrick, Kay},
  journal={Communications on Pure and Applied Mathematics},
  volume={65},
  number={5},
  pages={727--757},
  year={2012},
  publisher={Wiley Online Library}
}

@article{rumpf_simple_2004,
	title = {Simple statistical explanation for the localization of energy in nonlinear lattices with two conserved quantities},
	volume = {69},
	issn = {1539-3755, 1550-2376},
	url = {https://link.aps.org/doi/10.1103/PhysRevE.69.016618},
	doi = {10.1103/PhysRevE.69.016618},
	urldate = {2023-09-20},
	journal = {Phys. Rev. E},
	author = {B. Rumpf},
	year = {2004},
	pages = {016618},
}

@article{RUMPF20092067,
title = {Stable and metastable states and the formation and destruction of breathers in the discrete nonlinear {Schr{\"o}dinger} equation},
journal = {Physica D: Nonlinear Phenomena},
volume = {238},
number = {20},
pages = {2067-2077},
year = {2009},
issn = {0167-2789},
doi = {https://doi.org/10.1016/j.physd.2009.08.006},
url = {https://www.sciencedirect.com/science/article/pii/S0167278909002541},
author = {B. Rumpf}
}

@article{EBRAHIMI2025134905,
title = {Dynamics of localised states in the stochastic discrete nonlinear {Schr\"odinger} equation},
journal = {Physica D},
volume = {482},
pages = {134905},
year = {2025},
issn = {0167-2789},
doi = {https://doi.org/10.1016/j.physd.2025.134905},
url = {https://www.sciencedirect.com/science/article/pii/S0167278925003823},
author = {M. Ebrahimi and B. Drossel and W. Just}
}

@article{YOSHIDA1990262,
title = {Construction of higher order symplectic integrators},
journal = {Phys. Lett. A},
volume = {150},
number = {5},
pages = {262-268},
year = {1990},
issn = {0375-9601},
doi = {https://doi.org/10.1016/0375-9601(90)90092-3},
url = {https://www.sciencedirect.com/science/article/pii/0375960190900923},
author = {H. Yoshida}
}

@article{rasmussen_statistical_2000,
        title = {Statistical {Mechanics} of a {Discrete} {Nonlinear} {System}},
        volume = {84},
        issn = {0031-9007, 1079-7114},
        url = {https://link.aps.org/doi/10.1103/PhysRevLett.84.3740},
        doi = {10.1103/PhysRevLett.84.3740},
        urldate = {2023-08-03},
        journal = {Phys. Rev. Lett.},
        author = {Rasmussen, K. {{\O}}. and Cretegny, T. and Kevrekidis, P. G. and Gr{{\o}}nbech-Jensen, Niels},
        year = {2000},
        pages = {3740--3743}
}

@article{rumpf_transition_2008,
        title = {Transition behavior of the discrete nonlinear {Schr\"odinger} equation},
        volume = {77},
        issn = {1539-3755, 1550-2376},
        url = {https://link.aps.org/doi/10.1103/PhysRevE.77.036606},
        doi = {10.1103/PhysRevE.77.036606},
        number = {3},
        urldate = {2023-08-03},
        journal = {Phys. Rev. E},
        author = {B. Rumpf},
        year = {2008},
        pages = {036606}
}

@article{BaIuLiVu_PR21,
        title = {Statistical mechanics of systems with negative temperature},
        volume = {923},
        doi = {10.1016/j.physrep.2021.03.007},
        journal = {Phys. Rep.},
        author = {M. Baldovin and S. Iubini and R. Livi and A. Vulpiani},
        year = {2021},
        pages = {1-50}
}

@article{IuLeLiPo_JSM13,
        title = {Langevin dynamics of the discrete nonlinear {Schr\"odinger} chain},
        volume = {2013},
	doi = {10.1088/1742-5468/2013/},
        journal = {J. Stat. Mech.},
        author = {S. Iubini and S. Lepri and R. Livi and A. Politi},
        year = {2013},
        pages = {P08017}
}

@article{IuPo_PRL25,
        title = {Effective grand canonical description of
condensation in negative-temperature regimes},
        volume = {134},
	doi = {10.1103/PhysRevLett.134.097102},
        journal = {Phys. Rev. Lett.},
        author = {S. Iubini and A. Politi},
        year = {2025},
        pages = {097102}
}

@article{HaOl_SPDE23,
        title = { A stochastic thermalization of the
Discrete Nonlinear {Schr\"odinger} Equation},
        volume = {134},
	doi = {10.1007/s40072-022-00263-9},
        journal = {Stoch. PDE: Anal. Comp.},
        author = {A. Hannani and S. Olla},
        year = {2023},
        pages = {1379-1415}
}

@article{BoDe_CMP99,
        title = {A {Stochastic} {Nonlinear} {Schr\"odinger} {Equation} with {Multiplicative} {Noise}},
	volume = {205},
        doi = {10.1007/s002200050672},
        number = {1},
        urldate = {2025-04-07},
        journal = {Comm. Math. Phys.},
        author = {A. {de Bouard} and A. Debussche},
        year = {1999},
        pages = {161--181},
}

@article{RuNe_PRL01,
  title = {Coherent Structures and Entropy in Constrained, Modulationally Unstable,
Nonintegrable Systems},
  author = {B. Rumpf and A.C. Newell},
  journal = {Phys. Rev. Lett.},
  volume = {87},
  pages = {054102},
  year = {2001},
  doi = {10.1103/PhysRevLett.87.054102}
}

@article{IuFrLiOpPo_NJP13,
doi = {10.1088/1367-2630/15/2/023032},
year = {2013},
volume = {15},
pages = {023032},
author = {S. Iubini and R. Franzosi and R. Livi and {G.-L.}Oppo and A. Politi},
title = {Discrete breathers and negative-temperature states},
journal = {New J. Phys.}
}

@article{MiKaDaFl_PRL18,
year = {2018},
volume = {120},
pages = {184101},
author = {T. Mithun and Y. Kati and C. Danieli and S. Flach},
title = {Weakly Nonergodic Dynamics in the {Gross-Pitaevskii} Lattice},
journal = {Phys. Rev. Lett.},
doi={10.1103/PhysRevLett.120.184101}
}

@article{PhysRevLett.122.084102,
  title = {Dynamical Freezing of Relaxation to Equilibrium},
  author = {S. Iubini and L. Chirondojan and {G.L.} Oppo and A. Politi, Antonio and P. Politi},
  journal = {Phys. Rev. Lett.},
  volume = {122},
  pages = {084102},
  year = {2019},
  doi = {10.1103/PhysRevLett.122.084102}
}

@article{gradenigo2021condensation,
  title={Condensation transition and ensemble inequivalence in the discrete nonlinear Schr{\"o}dinger equation},
  author={Gradenigo, Giacomo and Iubini, Stefano and Livi, Roberto and Majumdar, Satya N},
  journal={The European Physical Journal E},
  volume={44},
  number={3},
  pages={29},
  year={2021},
  publisher={Springer}
}

@article{GrMa_JSM19,
  title = {A ﬁrst-order dynamical transition in the displacement distribution of a driven run-and-tumble particle},
  author = {G. Gradenigo and {S. N.} Majumdar},
  journal = {J. Stat Mech.},
  volume = {2019},
  pages = {053206},
  year = {2019},
  doi = {10.1088/1742-5468/ab11be}
}

@article{GrIuLiMa_JSM21,
  title = {Localization transition in the discrete nonlinear {Schr\"odinger} equation:
ensembles inequivalence and negative temperatures},
  author = {G. Gradenigo1 and S. Iubini and R. Livi and {S. N.} Majumdar},
  journal = {J. Stat Mech.},
  volume = {2021},
  pages = {023201},
  year = {2021},
  doi = {10.1088/1742-5468/abda26}
}

@article{giachello2025localization,
  title={Localization and entanglement entropy in the Discrete Non-Linear Schr{\"o}dinger Equation},
  author={Giachello, Martina and Iubini, Stefano and Livi, Roberto and Gradenigo, Giacomo},
  journal={arXiv preprint arXiv:2503.14364},
  year={2025}
}

@article{FoKaMa_JMP65,
  title = {Statistical Mechanics of Assemblies of Coupled Oscillators},
  author = {{G. W.} Ford and M. Kac and P. Mazur},
  journal = {J. Math. Phys.},
  volume = {6},
  pages = {504-515},
  year = {1965},
  doi = {10.1063/1.1704304}
}

@article{Zwan_JCP60,
 AUTHOR = {R.~Zwanzig},
 TITLE = {Ensemble method in the theory of irreversibility},
 JOURNAL = {J.\ Chem.\ Phys.},
 YEAR = 1960,
 VOLUME = 33,
 doi ={10.1063/1.1731409},
 PAGES =  1338}

@article{Naka_PTP58,
 AUTHOR = {S.~Nakajima},
 TITLE = {On Quantum Theory of Transport Phenomena},
 JOURNAL = {Prog.\ Theor.\ Phys.},
 YEAR = 1958,
 VOLUME = 20,
 doi = {10.1143/PTP.20.948},
 PAGES =  948}

@book{FiSa:90,
 AUTHOR= {E.~Fick and G.~Sauermann},
 TITLE={The quantum statistics of dynamic processes},
 PUBLISHER= {Springer},
 ADDRESS= {Berlin},
 YEAR= 1990}

@article{HiMo_PRL09,
 AUTHOR = {{M- C.} Hickey and {J. S.} Moodera},
 TITLE = {Origin of Intrinsic Gilbert Damping},
 JOURNAL = {Phys. Rev. Lett.},
 YEAR = 2009,
 VOLUME = 102,
 doi = {10.1103/PhysRevLett.102.137601},
 PAGES =  137601}

@book{Risk:89,
 AUTHOR="H.~Risken",
 TITLE={The Fokker-Planck equation : methods of solution
        and applications},
 PUBLISHER="Springer",
 ADDRESS= {Berlin},
 YEAR= 1989}

@article{ElFi_PA83,
 AUTHOR = {{F. J.} Elmer and E. Fick},
 TITLE = {Strongly driven disspative systems - an approach by {Robertson's} theory},
 JOURNAL = {Physica A},
 YEAR = 1983,
 VOLUME = 117,
 doi = {10.1016/0378-4371(83)90033-X},
 PAGES =  {232-242}}

@article{Kiku_JAP56,
 AUTHOR = {R. Kikuchi},
 TITLE = {On the Minimum of Magnetization Reversal Time},
 JOURNAL = {J. Appl. Phys.},
 YEAR = 1956,
 VOLUME = 27,
 doi = {10.1063/1.1722262},
 PAGES =  {1352-1357}}

@article{rumpf_growth_2007,
	title = {Growth and erosion of a discrete breather interacting with {Rayleigh}-{Jeans} distributed phonons},
	volume = {78},
	url = {https://doi.org/10.1209/0295-5075/78/26001},
	doi = {10.1209/0295-5075/78/26001},
	number = {2},
	journal = {Europhysics Letters},
	author = {Rumpf, B.},
	month = mar,
	year = {2007},
	pages = {26001},
}

@article{PhysRev.103.20,
  title = {Thermodynamics and Statistical Mechanics at Negative Absolute Temperatures},
  author = {Ramsey, Norman F.},
  journal = {Phys. Rev.},
  volume = {103},
  issue = {1},
  pages = {20--28},
  numpages = {0},
  year = {1956},
  month = Jul,
  publisher = {American Physical Society},
  doi = {10.1103/PhysRev.103.20},
  url = {https://link.aps.org/doi/10.1103/PhysRev.103.20}
}

@article{PhysRevLett.134.097102,
  title = {Effective Grand Canonical Description of Condensation in Negative-Temperature Regimes},
  author = {Iubini, Stefano and Politi, Antonio},
  journal = {Phys. Rev. Lett.},
  volume = {134},
  issue = {9},
  pages = {097102},
  numpages = {6},
  year = {2025},
  month = Mar,
  publisher = {American Physical Society},
  doi = {10.1103/PhysRevLett.134.097102},
  url = {https://link.aps.org/doi/10.1103/PhysRevLett.134.097102}
}

@ARTICLE{1158804,
  author={Sukhorukov, A.A. and Kivshar, Y.S. and Eisenberg, H.S. and Silberberg, Y.},
  journal={IEEE Journal of Quantum Electronics}, 
  title={Spatial optical solitons in waveguide arrays}, 
  year={2003},
  volume={39},
  number={1},
  pages={31-50},
  doi={10.1109/JQE.2002.806184}}

@article{b3727bba76384918a1c381ef76df660d,
title = "Experimental observation of linear and nonlinear optical bloch oscillations",
author = "R Morandotti and U Peschel and JS Aitchison and HS Eisenberg and K Silberberg",
year = "1999",
month = jan,
day = "1",
doi = "10.1103/PhysRevLett.83.4756",
volume = "83",
pages = "4756--4759",
journal = "Physical review letters",
issn = "0031-9007",
publisher = "American Physical Society",
number = "23",

}

@article{PhysRevLett.82.3288,
  title = {Observation of Intrinsically Localized Modes in a Discrete Low-Dimensional Material},
  author = {Swanson, B. I. and Brozik, J. A. and Love, S. P. and Strouse, G. F. and Shreve, A. P. and Bishop, A. R. and Wang, W.-Z. and Salkola, M. I.},
  journal = {Phys. Rev. Lett.},
  volume = {82},
  issue = {16},
  pages = {3288--3291},
  numpages = {0},
  year = {1999},
  month = Apr,
  publisher = {American Physical Society},
  doi = {10.1103/PhysRevLett.82.3288},
  url = {https://link.aps.org/doi/10.1103/PhysRevLett.82.3288}
}

@article{science.1062612,
author = {F. S. Cataliotti  and S. Burger  and C. Fort  and P. Maddaloni  and F. Minardi  and A. Trombettoni  and A. Smerzi  and M. Inguscio },
title = {Josephson Junction Arrays with Bose-Einstein Condensates},
journal = {Science},
volume = {293},
number = {5531},
pages = {843-846},
year = {2001},
doi = {10.1126/science.1062612},
URL = {https://www.science.org/doi/abs/10.1126/science.1062612},
eprint = {https://www.science.org/doi/pdf/10.1126/science.1062612}}

@article{Ng_2009,
doi = {10.1088/1367-2630/11/7/073045},
url = {https://doi.org/10.1088/1367-2630/11/7/073045},
year = {2009},
month = jul,
publisher = {},
volume = {11},
number = {7},
pages = {073045},
author = {Ng, G S and Hennig, H and Fleischmann, R and Kottos, T and Geisel, T},
title = {Avalanches of Bose–Einstein condensates in leaking optical lattices},
journal = {New Journal of Physics}
}

@Article{e19090445,
AUTHOR = {Iubini, Stefano and Lepri, Stefano and Livi, Roberto and Oppo, Gian-Luca and Politi, Antonio},
TITLE = {A Chain, a Bath, a Sink, and a Wall},
JOURNAL = {Entropy},
VOLUME = {19},
YEAR = {2017},
NUMBER = {9},
ARTICLE-NUMBER = {445},
URL = {https://www.mdpi.com/1099-4300/19/9/445},
ISSN = {1099-4300},
DOI = {10.3390/e19090445}
}

@article{Iubini_2017,
doi = {10.1088/1742-5468/aa7871},
url = {https://doi.org/10.1088/1742-5468/aa7871},
year = {2017},
month = jul,
publisher = {IOP Publishing and SISSA},
volume = {2017},
number = {7},
pages = {073201},
author = {Iubini, Stefano and Politi, Antonio and Politi, Paolo},
title = {Relaxation and coarsening of weakly-interacting breathers in a simplified DNLS chain},
journal = {Journal of Statistical Mechanics: Theory and Experiment}
}

@article{Iubini_2013,
doi = {10.1088/1367-2630/15/2/023032},
url = {https://doi.org/10.1088/1367-2630/15/2/023032},
year = {2013},
month = feb,
publisher = {IOP Publishing},
volume = {15},
number = {2},
pages = {023032},
author = {Iubini, S and Franzosi, R and Livi, R and Oppo, G-L and Politi, A},
title = {Discrete breathers and negative-temperature states},
journal = {New Journal of Physics}
}

@article{PhysRevLett.86.2353,
  title = {Discrete Solitons and Breathers with Dilute Bose-Einstein Condensates},
  author = {Trombettoni, Andrea and Smerzi, Augusto},
  journal = {Phys. Rev. Lett.},
  volume = {86},
  issue = {11},
  pages = {2353--2356},
  numpages = {0},
  year = {2001},
  month = Mar,
  publisher = {American Physical Society},
  doi = {10.1103/PhysRevLett.86.2353},
  url = {https://link.aps.org/doi/10.1103/PhysRevLett.86.2353}
}

@misc{giusfredi_mean-field_2025,
	title = {Mean-field theory of the {DNLS} equation at positive and negative absolute temperatures},
	url = {http://arxiv.org/abs/2511.09206},
	doi = {10.48550/arXiv.2511.09206},
	urldate = {2025-11-17},
	publisher = {arXiv},
	author = {Giusfredi, Michele and Iubini, Stefano and Politi, Antonio and Politi, Paolo},
	month = nov,
	year = {2025},
	note = {arXiv:2511.09206 [cond-mat]},
}

@article{Bray_AP93,
        title = {Theory of phase-ordering kinetics},
        volume = {43},
        doi = {10.1080/00018730110117433},
        journal = {Adv. Phys.},
        author = {A.J. Bray},
        year = {2002},
        pages = {357-459}
}

@book{HoLe:84,
 AUTHOR= {W. Horsthemke and R. Lefever},
 TITLE={Noise-induced transitions},
 PUBLISHER= {Springer},
 ADDRESS= {Berlin},
 doi= {10.1007/3-540-36852-3},
 YEAR= 1983}

@article{FLACH1998181,
title = {Discrete breathers},
journal = {Physics Reports},
volume = {295},
number = {5},
pages = {181-264},
year = {1998},
issn = {0370-1573},
doi = {https://doi.org/10.1016/S0370-1573(97)00068-9},
url = {https://www.sciencedirect.com/science/article/pii/S0370157397000689},
author = {S. Flach and C.R. Willis}
}

\end{document}